\documentclass[12pt,]{article}
  \usepackage{graphicx}
 \usepackage[cp1251]{inputenc}  
 \usepackage{indentfirst}       

\begin{document}

 \centerline {\large\bf Subdwarfs and white dwarfs from the {\it 2MASS}, {\it
 Tycho-2},}
 \centerline {\large\bf {\it XPM} and {\it UCAC3} catalogues}
 \bigskip
 \centerline {George~A.~Gontcharov,$^1$ Anisa~T.~Bajkova,$^1$}
 \centerline {Peter~N.~Fedorov,$^2$ and Vladimir~S.~Akhmetov$^2$}

 \bigskip
 \centerline {\small
  $^1$Central astronomical observatory at Pulkovo of RAS,}
 \centerline {\small Pulkovskoye chaussee 65/1, 196140, Saint-Petersburg, Russia.}
 \centerline {\small  $^2$Institute of Astronomy of Kharkiv National
 University,}
 \centerline {\small  Sums'ka 35, 61022 Kharkiv, Ukraine }


\begin{abstract}

The photometry from 2MASS, UCAC3 and SuperCosmos catalogues
together with the proper motions from the Tycho-2, XPM and UCAC3
catalogues are used to select the all-sky samples of 34 white
dwarfs, 1996 evolved and 7769 unevolved subdwarfs candidates for R
from 9 to 17 magnitude. The samples are separated from the main
sequence with admixture less than 10\% owing to the detailed
analysis of the distribution of the stars in the different color
index (CI) vs. reduced proper motion (RPM) diagrams for various
latitudes with special attention to the estimation of admixtures
in the samples and with Monte-Carlo simulation. It is shown that
the XPM and UCAC3 have the same level of proper motion accuracy.
Most of the selected stars has at least 6-band photometry with
accuracy is proved to be better than 0.2. The multi-color
photometry allows us to eliminate some admixtures and reveal some
binaries. The empirical calibrations of absolute magnitude versus
CI and RPM for Hipparcos stars give us photometric distances and
3D distribution for all the stars. Although the selection method
and uneven distribution of the XPM and UCAC3 data provide
noticeable biases the preliminary conclusions are made. The
subdwarfs show some concentration to the galactic centre
hemisphere with voids because of extinction in the Gould belt and
galactic plane. Some yet unexplained overdensities of the evolved
subdwarfs are seen in several parts of the sky. For 183 stars with
radial velocities 3D motion and galactic orbits are calculated.
For 56 stars with Fe/H collected from various sources we find the
relations of the metallicity with CI, asymmetric drift velocity
and orbital eccentricity. It is shown that most unevolved
subdwarfs belong to the halo with the scale height of $8\pm1$ kpc
and local mass density of halo subdwarfs of $2\cdot 10^{-5}
M_\odot~pc^{-3}$. Most evolved subdwarfs belong to the thick disk
with the scale height of $1.25\pm0.1$ kpc. Main parameters of the
selected stars are compiled into new SDWD catalogue for future
investigations. Special attention should be paid to spectroscopic
observations of these stars because 53\% of the selected white
dwarfs, 94\% of evolved and 98\% of unevolved subdwarfs are now
classified for the first time whereas the existed spectral
classification is wrong in many cases.

\end{abstract}

\section{Introduction}

Distribution, kinematics, age-velocity (AVR) and age-metallicity (AMR) relations of
the subluminous stars, such as subdwarfs (SD) and white dwarfs (WD), are important
yet poorly known data to study the structure, formation and history of the Galaxy.
Unfortunately, a hard work of spectroscopic detections, classifications and measurements
of the stars have been made for only hundreds of them at best and naturally sometimes are
erroneous.

New all-sky astrometric and photometric surveys of millions stars,
such as Tycho-2 \cite{tycho2}, UCAC3 \cite{ucac3}, 2MASS
\cite{2mass}, XPM \cite{xpm}, become a good source to select large
samples of the subluminous stars. This new
spectroscopy-independent look on the stars is possible by use of
multi-color photometry and its combination with proper motions
known as {\em the reduced proper motions} (RPM). It is determined
for the 2MASS Ks photometric band (and similar to others) as
\begin{equation}
M'_{Ks}=Ks+5+5\lg(\mu)
\label{mks}
\end{equation}
where $\mu$ is a total proper motion in arcsec~yr$^{-1}$.
The interstellar extinction $A_{Ks}$ should be added to the right part of this equation
to make $M'_{Ks}$ being an analog of the absolute magnitude $M_{Ks}$.
But an advantage of the use of the 2MASS
infra-red (IR) photometry is negligible value of $A_{Ks}$ for the nearest part of the Galaxy:
$A_{Ks}\approx0.1A_{V}$. Consequently, 2MASS been the only current all-sky source of the precise
IR photometry is ultimate for such investigations.

As shown by an intensive Monte-Carlo simulation \cite{model}
(hereafter GG2009), $M'_{Ks}$ can be used instead of absolute
magnitude $M_{Ks}$ to select some classes of stars and calculate
their {\em photoastrometric} distances via empirical calibration
\begin{equation}
M'_{Ks}~ vs.~ M_{Ks}
\label{mksmks}
\end{equation}
which is comparable to their {\em photometric} distances via calibration
\begin{equation}
(J-Ks)~ vs.~ M_{Ks}
\label{jkmks}
\end{equation}
Then one can consider stellar 3D distribution and kinematics.
GG2009 proves that the introduced biases can be taken into the account by Monte-Carlo simulation.

The first investigation of SDs by use of the RPM was made by
\cite{jones}. Some approaches in the current paper are similar to
the ones in \cite{rcg} where the RPM were used for the selection
and study of clump red giants. The selection of a admixture-free
sample of SDs from the Tycho-2 by RPM and a color index was
proposed by GG2009. A noticeable investigation of the SDs from the
SDSS survey \cite{sdss} in {\em a part of the sky} is made by
\cite{smith}.

As discussed later, the white dwarfs are a by-product of this selection of the SDs
because sometimes these two classes could not be easy separated. Thus, we put more attention to
the SDs.

\section{Subdwarfs}

There are several definitions of SD depending mainly on the research method.
From the stellar evolution point of view they are the stars slightly hotter (bluer) than
solar metallicity zero-age main sequence (ZAMS) or solar metallicity zero-age horizontal branch
(ZAHB) stars of the same mass because of their lower-than-solar metallicity.
From the spectroscopic point of view they are the stars slightly fainter than
ZAMS and ZAHB of the same spectral class.
From the photometric point of view they are the stars in the well-defined respectively
unpopulated domain in a color-magnitude diagram.
From the kinematic point of view they are the stars with large motion with respect to the Sun
because all or majority of them are believed to belong to the galactic population II.
One could show that no way to unite all these definitions, first of all, because new
investigations reveal some stars that fit only part of these criteria.
Particularly, as shown later, not all SDs have low metallicity and population II kinematics.

We look at the SDs from the photometric-kinematic point of view selecting them
as all the stars in a domain on color-RPM ``$(J-Ks)$ -- $M'_{Ks}$'' diagram
(in addition to the color-absolute magnitude ``$(J-Ks)$ -- $M_{Ks}$'' diagram for some of the SDs).
Therefore, we do not pay attention to the classic spectroscopic classification of SDs
into sdO, sdB, etc..
Consequently, the selected samples can be considered as SD and WD {\em candidates}
for spectroscopy. But even without spectroscopy this method allows us to consider the distribution
of the samples on colors, spectral classes, velocities, metallicities, ages, galactic populations,
and so on.

\begin{figure}
\begin{center}
\includegraphics[width = 140mm]   {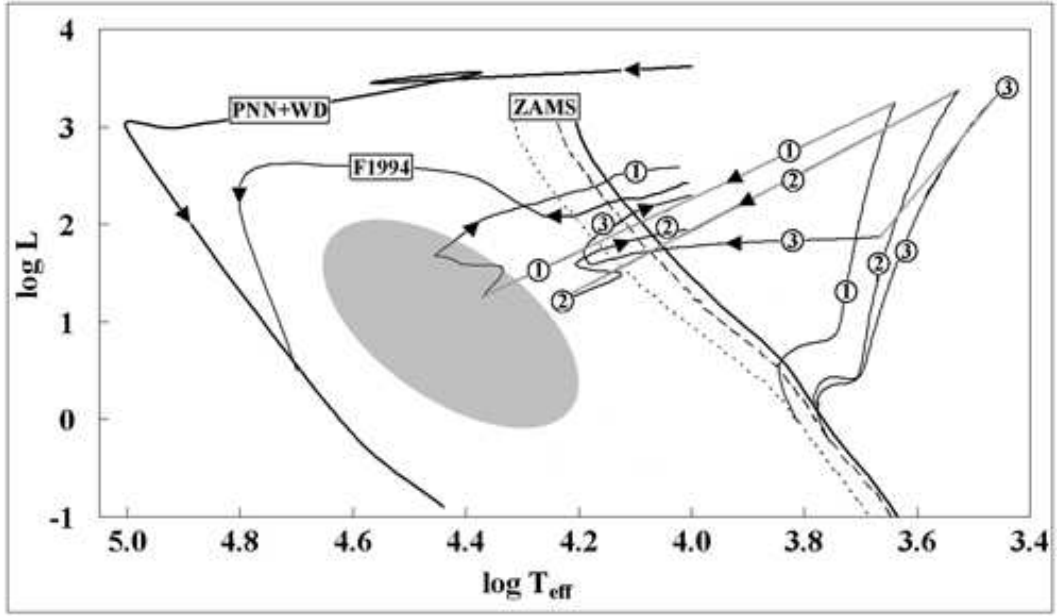}
\caption{$\lg~T_{eff}$ vs. $\lg~L$ for ZAMS with $Z=0.019$, 0.008
and 0.0004 --- thick, dashed and dotted lines marked as ZAMS;
PNN+WD track for $0.9M_\odot$, $Z=0.004$ --- thick line at left
marked as PNN+WD; PNN+WD track from \cite{fbbc} for $0.5M_\odot$,
$Z=0.05$, $Y=0.35$ marked as F1994; and 3 evolutionary tracks for
stars of $0.9M_\odot$ before and $0.5M_\odot$ after helium flash,
marked as 1, 2, 3 for $Z=0.0004$ with $Y=0.23$, $Z=0.008$ with
$Y=0.25$ and $Z=0.04$ with $Y=0.46$ respectively. The grey ellipse
shows the domain of subdwarfs generated after evolution in close
binaries.}
\end{center}
\end{figure}

The position of some SDs in the effective temperature - luminosity
(``$\lg~T_{eff}$ -- $\lg~L$'') diagram is shown in Fig.~1. The
solar metallicity ZAMS is shown as a thick line as well as the
ones for $Z=0.008$ and $Z=0.0004$ are shown as dashed and dotted
lines respectively and marked as ``ZAMS'' \cite{g2000}. As an
example, three evolutionary tracks are shown by numbered thin
lines (with jumps from the helium flash to ZAHB shown by grey):
\begin{enumerate}
\item $Z=0.0004$, $Y=0.23$, $M=0.9M_\odot$ before and
$M=0.5M_\odot$ after helium flash \cite{g2000}, \item $Z=0.008$,
$Y=0.25$ and the same masses \cite{g2000}, \item $Z=0.04$,
$Y=0.46$ and the same masses \cite{bgmn}.
\end{enumerate}
Such tracks crossing solar ZAMS twice (deviating from ZAMS and
ZAHB) are possible only in models with convective overshooting and
never exist in the classical ones. Some models allow such
considerable mass loss from $0.9M_\odot$ to $0.5M_\odot$ at the
red giant branch (RGB) and its tip with helium flash \cite{cssw}.
But even a star with less mass loss, from $0.7M_\odot$ to
$0.5M_\odot$ can be a SD \cite{g2000}. A classic stellar
evolutionary track for planetary nebula nucleus (PNN) stage and
white dwarf (WD) sequence with helium burning, $M=0.9M_\odot$,
$Z=0.004$ is also shown as a thick line at high $T_{eff}$ marked
as ``PNN+WD'' \cite{vw}. Another PNN-WD track from \cite{fbbc} for
$M=0.5M_\odot$, $Z=0.05$, $Y=0.35$ is shown as the thin line
marked as ``F1994''. It means that low mass PNN and WD are at the
edge inside the SD domain and, hence, sometimes they may be
confused with SD. The grey ellipse shows the domain of the SDs
after evolution in close binaries \cite{han}. It expands the SD
domain considerably: apparently SD may be almost anywhere on the
``$\lg~T_{eff}$ -- $\lg~L$'' plane between the solar ZAMS and the
classic PNN-WD track (2 thick lines in the Fig.~1). But single
(and without a binary past) very hot SDs have not been explained
by known evolutionary tracks.

Two SD domains are seen {\em from the models} fainter and hotter than the solar ZAMS:
\begin{description}
\item[evolved SDs] with $4.05<\lg~T_{eff}<4.8$ ($11000<T_{eff}<60000$ K, spectral
classes sdO and sdB, $(J-Ks)<0$, $0<\lg~L<3$, $2<M_{Ks}<8$, hereafter ESDs),
\item[unevolved SDs] with $\lg~T_{eff}<3.85$ ($T_{eff}<7000$ K, spectral
classes sdF and later, $(J-Ks)>0.1$, $\lg~L<0.5$, $M_{Ks}>2$, hereafter USDs).
\end{description}
The gap between these domains should not be populated by single USDs because they would have
low metallicity but been born within the last 7 Gyr.
However, some binaries of a ESD and a redder dwarf must have $(J-Ks)>0$ and fill the gap.
Real data should be used to test it.

To be rather numerous in the large surveys the SDs
1) stay in the domain for a long time: several Gyr for USD and about 100 Myr for ESD,
2) deviate considerably from the solar ZAMS (the data used must be quite precise).

The USDs look as a quite homogeneous population: low-metallicity
population II stars with little or no galactic rotation. By
definition they are at a low-metallicity ZAMS. Therefore, their
distribution on mass and age is determined by yet poorly known
birthrate. For example, if all low-metallicity stars were born
only more than 7 Gyr ago then all USDs have masses $M<0.9M_\odot$
(more massive SDs have leaved ZAMS) and, hence, $T_{eff}$, spectra
and $(J-Ks)$ are as pointed out early. However, if some
low-metallicity stars were born within the last 7 Gyr in a
satellite merged by the Galaxy then one could find more massive,
hot and bright USD.

In contrast to USDs the ESDs is a heterogeneous population. The
only common feature for ESDs is the mass nearly 0.4-0.6M$_\odot$
because no a theoretical track of a more massive star in the ESD
domain but less massive ones are not so evolved. Many of ESDs are
helium rich and the most of them are helium core burning stars
with extremely thin hydrogen envelops (the extended horizontal
branch (EHB) stars). The tracks in Fig.~1 show that ESDs have
various metallicities: very low, nearly solar, very high and even
unusual ones with low Z and high Y.

Moreover, ESDs look even more complex when one takes into the
account other ways putting the evolved stars to the ESD domain.
The reason of this diversity of ESDs is that the usual
evolutionary way from the RGB through horizontal branch (HB),
asymptotic giant branch (AGB), PNN to WD sequence can be deviated
or interrupted to put a star into the ESD domain. Since new
evolutionary ways to ESD could be found in the future, we discuss
the known ones only briefly. Various scenarios of the mass loss at
the RGB and helium flash at the RGB tip, as well as the evolution
of close binaries are most important because future ESD has to be
quite massive before the helium flash to rich it for a reasonable
time yet much less massive to enter the ESD domain. A star becomes
a ESD coming from RGB tip \cite{catelan}, EHB \cite{fbbc}, the
early-AGB and AGB similar to PNN-WD way but inside the ESD domain
\cite{fbbc}, \cite{catelan}, WD (hot-flasher scenario with helium
flash delay by \cite{mb}), a pair of merging WDs or evolution of
binary \cite{han}, and probably some other exotic ways. Thus, the
ESDs can have various ages, metallicities as well as various
velocities with respect to the galactic centre which is known as
the asymmetric drift of the different populations at the solar
distance from the centre.

\section{The data}

The Tycho-2 $B_{T}$, $V_{T}$ photometry and proper motions are used when the precision is
better than $0.2^{m}$ and 7 milliarcsec~year$^{-1}$ (hereafter mas~yr$^{-1}$) respectively.

The XPM catalogue is made in Kharkov National University, Ukraine.
It combines the positions from the 2MASS and USNO-A2.0
\cite{usnoa2} catalogues in order to derive the absolute proper
motions of about 280 million stars distributed all over the sky
excluding a small region near the Galactic Centre, in the
magnitude range $12^{m}<B<19^{m}$. The mean epoch difference of
the positions used is about of 45 years for the Northern
hemisphere and 17 years for the Southern one. The zero-point of
the absolute proper motion frame (the ``absolute calibration'')
was specified with the use of about 1.45 million galaxies from the
2MASS. Most of the systematic zonal errors inherent in the
USNO-A2.0 catalogue were eliminated before the calculation of
proper motions. The mean formal error of absolute calibration is
less than 1 mas $yr^{-1}$ \cite{xpm}.

The third U.S. Naval Observatory CCD Astrograph Catalog, UCAC3 is
a compiled, all-sky star catalogue covering mainly the 8 to 16
magnitude range in a single bandpass between V and R. We use its
proper motions and photometry. The latter are UCAC3 own band
photometry (hereafter $R_{UCAC3}$), 3 bands of 2MASS (J, H, Ks)
and 3 bands from the SuperCosmos project \cite{sc} (hereafter
$B_{SC}$, $R_{SC}$, $I_{SC}$). The proper motions of bright stars
are based on about 140 catalogs, including Hipparcos \cite{hip},
Tycho and all catalogs used for the Tycho-2 proper motion
construction. Proper motions of faint stars are based on a
re-reduction of early epoch SPM data ($\delta$ from $-90^\circ$ to
$-10^\circ$) plus Schmidt plate data from the SuperCosmos project
(down weighted due to systematic errors of order 0.1 arcsec).  The
proper motions of faint stars ($R_{UCAC3}\geq$13.5) therefore
should be used with caution \cite{ucac3}. The unpublished plate
measure data from the several astrometric catalogues have
considerably contributed to improve proper motions for stars
mainly in the 10 to 14 mag range (the interval which is most
interesting for us); however, these data do not cover all sky as
pointed out by \cite{ucac3}.

As pointed out early, precise IR photometry is the key data for
our study. Therefore, we consider only stars with $6<Ks<14$
following the error budget of the 2MASS photometry. The UCAC3 and
SuperCosmos photometry cover all this photometric interval but not
for all stars. The $R_{UCAC3}$ accuracy is believed to be at the
level of 0.1-0.2 mag whereas SuperCosmos typical photometric
accuracy is about of 0.3 mag. However, as declared by \cite{sc},
the $B_{SC}-R_{SC}$ and $R_{SC}-I_{SC}$ colors are accurate to
0.07 mag.

There are about 47 millions stars with $Ks<14$ common to the
2MASS, XPM and UCAC3. The mean difference of their proper motions
in the sense ``UCAC3 minus XPM'' is quite small:
$\overline{\Delta\mu_{\alpha}\cos(\delta)}=-2.8$ mas~yr$^{-1}$,
$\overline{\Delta\mu_{\delta}}=+0.3$ mas~yr$^{-1}$. The standard
deviations of their proper motion differences are
$\sigma(\Delta\mu_{\alpha}\cos(\delta))=12$ mas~yr$^{-1}$,
$\sigma(\Delta\mu_{\delta})=12$ mas~yr$^{-1}$. Taking into the
account that majority of the common stars has $Ks\approx14$ their
accuracy of the UCAC3 proper motion components is near 6-8
mas~yr$^{-1}$ as declared for faint stars by \cite{ucac3}.
Consequently, we conclude that the XPM proper motion accuracy has
nearly the same level. The detail comparison of the XPM and UCAC3
as well as analysis of the XPM proper motion errors are presented
elsewhere \cite{xpm1}.

The proper motion difference ``UCAC3 minus XPM'' seems not to depend on XPM epoch
difference or another parameter. Therefore, we simply select for the investigation
about 23.4 million stars with good precision of the UCAC3 proper motions and good
agreement of them with the ones from the XPM (in mas~yr$^{-1}$):
\begin{equation}
\Delta\mu_{\alpha}\cos(\delta)<10\;\mbox{, }\Delta\mu_{\delta}<10\;\mbox{, }
\sigma(\mu_{UCAC3})<15
\label{cond}
\end{equation}
The means of the UCAC3 and XPM proper motions of the stars are used hereafter,
except Tycho-2 stars among them for which the Hipparcos or Tycho-2 proper motions are used
depending on their formal errors.

\section{Selection of the stars}

\begin{figure}
\begin{center}
\includegraphics[width = 84mm]   {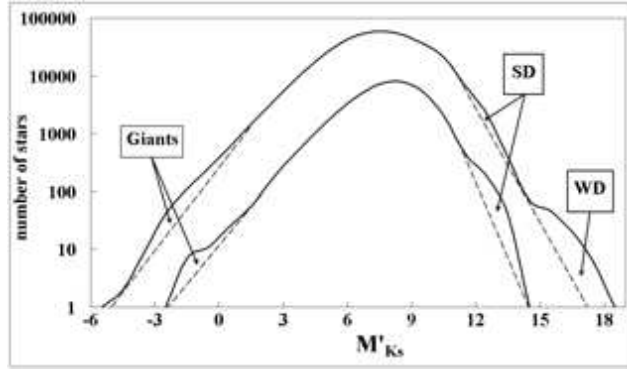}
\caption{The distribution of the XPM stars along $M'_{Ks}$ for
$0.34<(J-Ks)<0.35$ The higher thick curve is for $|b|<10^\circ$,
the lower thick one is for $|b|>50^\circ$. The deviations of these
curves from the dashed lines are giants, SDs and WDs.}
\end{center}
\end{figure}

In contrast to simple selections of SDs in the ``$(J-Ks)$ --
$M'_{Ks}$'' plane (for example, by \cite{smith}) we fulfill the
detailed analysis of the distribution of Tycho-2, XPM and UCAC3
SDs and WDs for the whole interval of $(J-Ks)$, $M'_{Ks}$ and all
latitudes as well as pay special attention to the estimation of
admixtures in the samples. In addition we analyze the distribution
of stars in the ``$(B_{T}-V_{T})$ -- $M'_{V_{T}}$'' and
``$(B_{SC}-I_{SC})$ -- $M'_{B_{SC}}$'' planes to eliminate
erroneous classifications due to duplicity (see later).

Firstly, the distribution of the stars in the ``$(J-Ks)$ --
$M'_{Ks}$'' plane is approximated by 4 Gaussians corresponding to
giants, main sequence (MS) stars, SDs and WDs. Then the MS-SD and
SD-WD cut lines are found between the Gaussians as some polynomial
functions of $(J-Ks)$ in order to make the samples containing no
more than 10\% of admixture stars. An example of the distribution
of the XPM stars along $M'_{Ks}$ for $0.34<(J-Ks)<0.35$ (a section
of 4 Gaussians) is presented in Fig.~2. The higher thick curve is
for $|b|<10^\circ$, where $b$ is the galactic latitude. The lower
thick curve is for $|b|>50^\circ$. The deviations of these curves
from the dashed lines are due to heterogeneous distribution of
stars on velocity (with respect to the Sun) and, hence, correspond
to non-MS stars, namely, giants, SDs and WDs. For
$0.34<(J-Ks)<0.35$ the SDs should be selected at
$12.25<M'_{Ks}<15$ and WDs at $M'_{Ks}>15$.

It is found that
\begin{itemize}
\item the distribution of Tycho-2 stars is strongly affected by
the magnitude limit of the catalogue about $V_{T}\approx11$ mag
and, hence, it contains few WDs and USDs; therefore, the cut lines
in the ``$(J-Ks)$ -- $M'_{Ks}$'' plane are taken from XPM and
UCAC3 only, \item the SD-WD cut line is affected by the magnitude
limits of the XPM and UCAC3 providing variations of SD admixture
in the WD sample for $M'_{Ks}\geq15$, therefore, this cut line is
not so well determined as the MS-SD one, \item SDs at different
$b$ cover the same interval of $M'_{Ks}$ (as seen in Fig.~2) and,
hence, the cut lines do not depend on $b$, \item the distributions
for UCAC3 and XPM are fairly the same that proves the same level
of accuracy of these catalogues, \item WDs are noticeable only at
low latitudes.
\end{itemize}
The final MS-SD cut line giving the SD sample with less than 10\% admixture is
\begin{equation}
M'_{Ks}>11.766(J-Ks)^{3}-16.804(J-Ks)^{2}+12.365(J-Ks)+9.5
\label{mssd}
\end{equation}
It is found that the SD-WD cut line has sense only for
$-0.1<(J-Ks)<0.7$ and for some latitudes. Star should be a WD if
\begin{equation}
M'_{Ks}>-13.036(J-Ks)^{2}+11.585(J-Ks)+12.65
\label{sdwd}
\end{equation}

A Monte-Carlo simulation similar to GG2009 is made to test the empirically obtained cut lines.
The content of 2MASS catalogue for MS, ESD, USD and WD stars is reproduced to analyze their
distribution in the ``$(J-Ks)$ -- $M'_{Ks}$'' plane. The photometric and astrometric errors with
realistic dispersions are applied.

We used normal and uniform distributions realized with the
Microsoft Excel 2007 random number generator, whose general
description was given by \cite{random}. We consider 4 categories
of stars: MS, ESD, USD and WD. For each category we generated
200000 model stars with specified
\begin{itemize}
\item the uniform distributions in rectangular galactic X and Y
coordinates and distribution in Z following \cite{robin} and
\cite{veltz}, \item the distributions in velocity components along
the galactic longitude $l$ and latitude $b$, $V_{l}$, and $V_{b}$
following \cite{robin} and \cite{veltz}, \item the distributions
in $(J-Ks)$, the dependence of $M_{Ks}$ on $(J-Ks)$ and its
scatter following \cite{g2000} and \cite{g2005} and variations of
the distribution of the stars on the metallicity, \item the
interstellar extinction $A_{Ks}$ calculated as $0.11A_{V}$ where
$A_{V}$ is calculated following new model of extinction by
\cite{gould} taking into the account the extinction in the Gould
belt, \item the errors of 2MASS photometry for $Ks<14$ accepted as
0.05 mag and errors of proper motions widely varied.
\end{itemize}

Then for every model star we calculate
\begin{itemize}
\item true distance $R=(X^2+Y^2+Z^2)^{1/2}$,
\item $l$ and $b$: $\tan(l)=Y/X$, $\tan(b)=Z/(X^2+Y^2)^{1/2}$,
\item $\mu_{l}=V_{l}/(4.74R)$, $\mu_{b}=V_{b}/(4.74R)$,
\item total proper motion $\mu=(\mu_{l}^2+\mu_{b}^2)^{1/2}$,
\item $Ks=M_{Ks}-5+5\lg(R)+A_{Ks}$,
\item truncation of the sample by $Ks<14$,
\item $M'_{Ks}$ from Ks, $\mu$ and $A_{Ks}$ following the equation~\ref{mks}.
\end{itemize}

The simulation proves that the input dispersions (including errors) translate into the output
dispersion of the stellar distribution on $M'_{Ks}$ so that
\begin{itemize}
\item the photometric errors at the level of 0.05 mag or lower do not influence the output
dispersions,
\item the reasonable variations of the metallicity distribution do not influence the output
dispersions,
\item the proper motion errors are very important: the MS output dispersion rises with it
so that {\em pure and complete} SD and WD samples for $(J-Ks)<0.7$ could be obtained
when the proper motion error is at the level of 1 mas~yr$^{-1}$ whereas
no SD or WD sample could be obtained when the error is at the level of 20 mas~yr$^{-1}$,
the level of the typical proper motion itself,
\item the obtained cut lines are correct to make 90\%-clear SD sample if the proper motion
mean error is about of 10 mas~yr$^{-1}$, but it gives strong biases because many slow stars
are lost,
\item no MS-SD-WD separation for $(J-Ks)>0.7$ (only $(J-Ks)<0.7$ are considered later).
\item the reddening because of extinction strongly affects the selection of ESDs:
the stars with large reddening are lost but the rest stars have negligible extinction,
\item as expected after GG2009, the purity, completeness in some region of space, and
symmetry (in favor to slow or fast stars) cannot be combined in the same sample,
and we prefer the purity accepting such cut lines.
\end{itemize}

\begin{figure}
\begin{center}
\includegraphics[width = 84mm]   {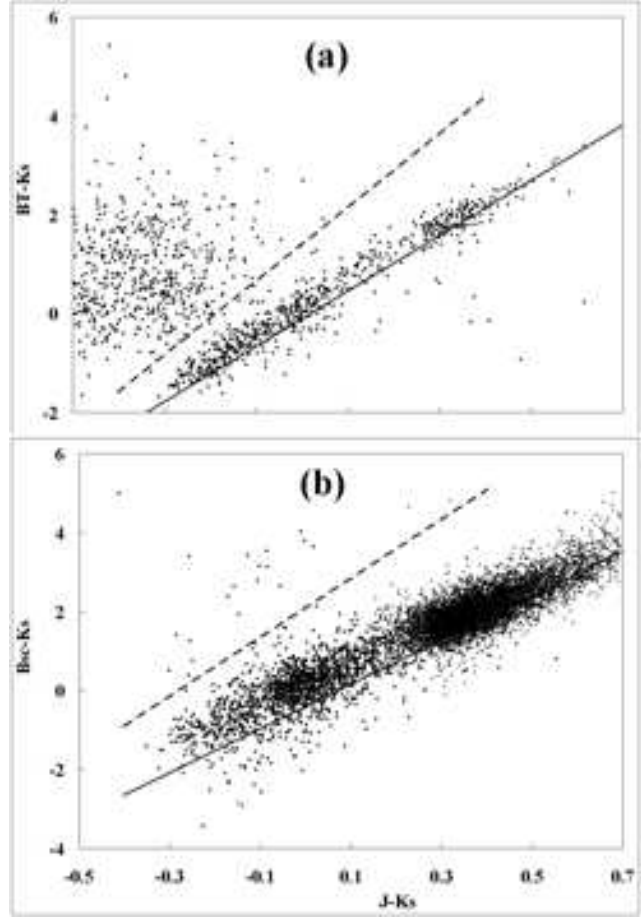}
\caption{The stars selected by use of the equation~\ref{mssd}: (a) Tycho-2 stars in the diagram
``$(J-Ks)$ vs. $(B_{T}-Ks)$'',
(b) XPM-UCAC3 ones in the diagram ``$(J-Ks)$ vs. $(B_{SC}-Ks)$''.
The empirical ZAMS are shown by solid lines.
The dashed lines show the theoretical reddening slope and separate the stars with normal
color relations from the ones with color discrepancies.}
\end{center}
\end{figure}

The stars selected by use of the equation~\ref{mssd} are shown in
Fig.~3: (a) Tycho-2 stars in the diagram ``$(J-Ks)$ vs.
$(B_{T}-Ks)$'' and (b) rest stars in the diagram ``$(J-Ks)$ vs.
$(B_{SC}-Ks)$''. The empirical ZAMS $(B_{T}-Ks)=5.6(J-Ks)-0.1$ and
the line $(B_{T}-Ks)=7.5(J-Ks)+1.4$ with theoretical reddening
slope both taken from \cite{ob} are shown in Fig.~3 (a) for
Tycho-2 stars by thick and dashed lines respectively. Similar
lines shifted due to $B_{T}-B_{SC}$ difference are shown in Fig.~3
(b) for rest stars: $(B_{SC}-Ks)=5.6(J-Ks)-0.4$ and
$(B_{SC}-Ks)=7.5(J-Ks)+2.1$. The spread of the dots near the ZAMS
is due to the photometric errors and reddening shift. The stars
lower than the ZAMS are prove to be binaries with composite
spectrum \cite{stark}. The stars higher than the reddening lines
show the discrepancy between their colors. It can be explain by
stellar duplicity, known in most cases or hidden. It gives
visual-IR identification mistakes and/or unusual energy
distribution in the common spectrum. However, one should not
simply eliminate these stars because some of them could contain SD
or WD components, specially, taking into account that the binary
fraction in ESD stars is much higher than for normal stars
\cite{osten}.

The pairs with subluminous components can be separated from the
rest binaries in the diagrams ``$(B_{T}-V_{T})$ -- $M'_{V_{T}}$''
and ``$(B_{SC}-I_{SC})$ -- $M'_{B_{SC}}$'' after the main
selection by use of the equation~\ref{mssd}. These two diagrams
for stars selected by use of the equation~\ref{mssd} are shown in
Fig.~4 (a) and (b) respectively. In fact, the former diagram
contains Tycho-2 stars and the latter one does the rest stars. The
Hipparcos stars \cite{hip2} with parallax relative error less than
0.3 are shown in the former diagram as the grey ``cloud'' of
points. It could not be provided for the latter diagram because no
precise $B_{SC}$ and $I_{SC}$ for the Hipparcos stars. The cut
lines are shown:  $M'_{V_{T}}=10(B_{T}-V_{T})+6$ and
$M'_{B_{SC}}=4.5(B_{SC}-I_{SC})+5$. All stars higher than the
lines are eliminated. The plane ``$(J-Ks)$ -- $M'_{Ks}$'' is
preferred for main selection because the $(J-Ks)$ for the stars is
much more precise than the $(B_{T}-V_{T})$ or $(B_{SC}-I_{SC})$.

\begin{figure}
\begin{center}
\includegraphics[width = 84mm]   {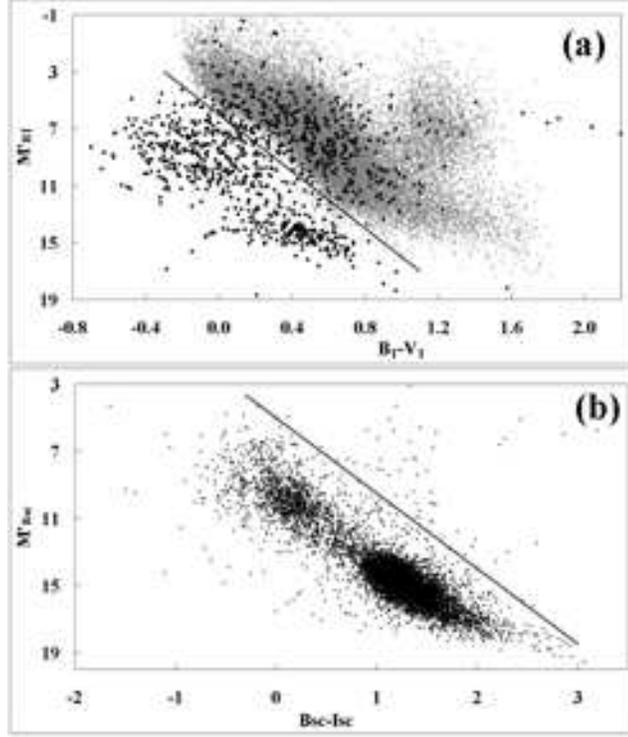}
\caption{The stars selected by use of the equation~\ref{mssd}: (a) Tycho-2 stars in the diagram
``$(B_{T}-V_{T})$ -- $M'_{V_{T}}$'',
(b) XPM-UCAC3 ones in the diagram ``$(B_{SC}-I_{SC})$ -- $M'_{B_{SC}}$''.
The lines separate the suspected SDs and WDs from the rest stars.}
\end{center}
\end{figure}

The final sample of 9799 SD and WD candidates is shown in Fig.~5. in the diagram
``$(J-Ks)$ -- $M'_{Ks}$'' together with the Hipparcos stars with parallax relative
error less than 0.3 (the grey ``cloud'' of points).
The 1040 stars selected from Tycho-2 are shown in subfigure (a) whereas 8759 ones from
XPM and UCAC3 are in subfigure (b) (358 Tycho-2 stars are also XPM and UCAC3 ones but their
Tycho-2 proper motions are preferred).
The MS-SD cut line is evident as the higher border of the black bulk.
The SD-WD cut line from the equation~\ref{sdwd} is shown as dashed line.
But it is obtained only for $(J-Ks)>-0.1$ and no WD found for $(J-Ks)>0.2$.
Therefore, for the selection of WDs we accept the polygonal thick line shown in the Fig.~5:
 $$
  M'_{Ks}>10.3-10(J-Ks),\qquad if (J-Ks)<-0.09;\\
 $$
 \begin{equation}
 M'_{Ks}>12.3+13.3(J-Ks),\qquad if (J-Ks)\ge-0.09.
 \label{poly}
\end{equation}
Since this SD-WD separation is a matter of convention, some our SDs with $(J-Ks)<-0.1$
may be WDs and vice versa.

The gap between the grey and black bulks means that we have to apply rather strong cut
to get rather pure sample. Therefore, we expect to lose
many or even majority of slow SDs and some known ones selected by spectroscopy.
However, the obtained large samples of stars with high probability to be SDs or WDs
seems to be useful not only for current study of their properties but for the next
spectroscopic studies.

\begin{figure}
\begin{center}
\includegraphics[width = 84mm]   {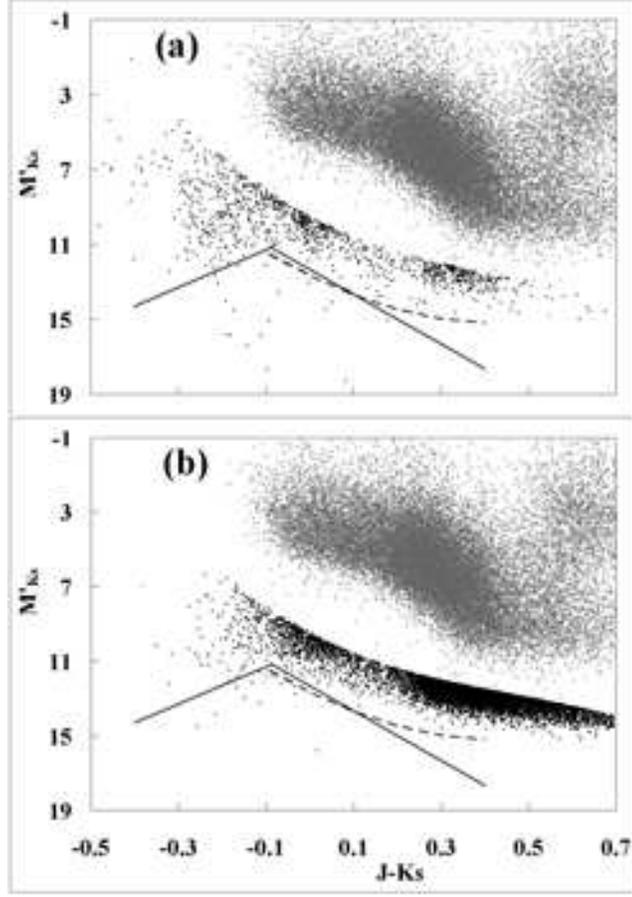}
\caption{Hipparcos stars with parallax relative error less than 0.3 (grey points)
together with selected (a) 1040 Tycho-2 and (b) 8759 XPM-UCAC3 stars (black diamonds),
SD-WD cut from equation~\ref{sdwd} (dashed line)
and accepted SD-WD cut (thick polygonal line).}
\end{center}
\end{figure}

Two domains, the ESDs and USDs are evident near $(J-Ks)\approx0$ and $\approx0.3-0.4$ respectively.
As expected the USDs are much more numerous than ESDs. But Tycho-2 contains too little
USDs because of the magnitude limitation.

The distribution of the Tycho-2 (open diamonds) and XPM-UCAC3 (crosses) selected stars
in the diagram ``$(J-Ks)$ vs. total proper motion'' is shown in Fig.~6.
The proper motions are in mas~yr$^{-1}$.
The both subsamples show ESD and USD overdensities at the same $(J-Ks)$.

The distribution of the selected stars on $(J-Ks)$ is shown in Fig.~7. The ESD and USD
Gaussians are evident here with some asymmetries due to binaries and reddening.

\begin{figure}
\begin{center}
\includegraphics[width = 84mm]   {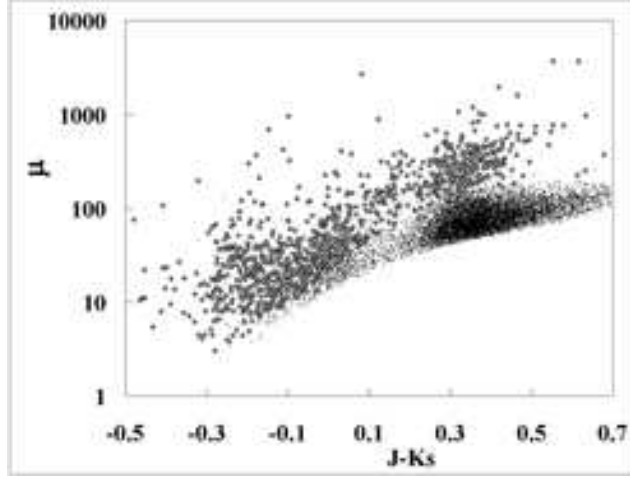}
\caption{The distribution of the Tycho-2 (open diamonds) and XPM-UCAC3 (crosses) selected stars
in the diagram ``$(J-Ks)$ vs. total proper motion (in mas~yr$^{-1}$)''.}
\end{center}
\end{figure}

\begin{figure}
\begin{center}
\includegraphics[width = 84mm]   {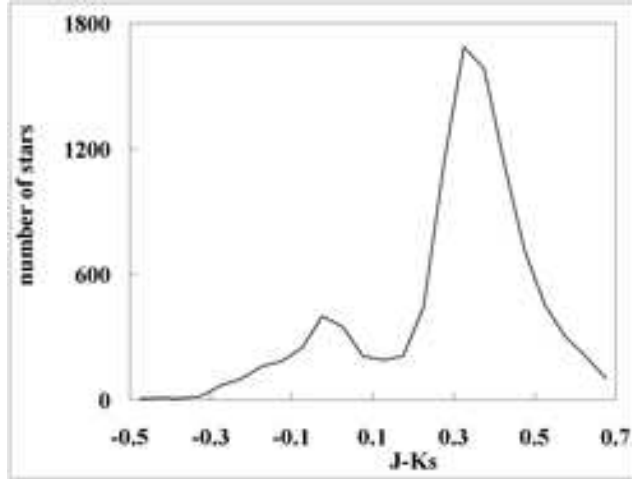}
\caption{The distribution of the selected stars on $(J-Ks)$.}
\end{center}
\end{figure}

The number of selected stars by band, magnitude range, median magnitude and median
photometric error are presented in Table~1.

\begin{table}
 \caption{Number of selected stars by band, magnitude range, median magnitude and median
photometric error.}
 \label{photo}
 \begin{tabular}{@{}lrrcc}
 \hline
  & stars & mag range & median mag & median error   \\
\hline
$B_{T}$    & 737  & 7.6-13.1 & 11.9 & 0.10 \\
$V_{T}$    & 648  & 7.3-12.8 & 11.7 & 0.13 \\
$B_{SC}$   & 8961 & 8.9-18.4 & 15.3 & $\approx$0.2\\
$R_{SC}$   & 9113 & 9.0-16.9 & 14.6 & $\approx$0.2\\
$I_{SC}$   & 9113 & 9.3-16.8 & 14.1 & $\approx$0.2\\
$R_{UCAC3}$ & 7434 & 9.4-17.3 & 15.0 & 0.13 \\
$J$        & 9799 & 6.7-14.7 & 13.8 & 0.03 \\
$Ks$       & 9799 & 6.5-14.0 & 13.5 & 0.03 \\
\hline
 \end{tabular}
 \medskip
\end{table}

The ESDs and WDs are separated from the USDs in some color-color
diagrams: examples of these separated overdensities are presented
in Fig.~8. The reddest overdensities are USDs, the middle and the
bluest ones are ESDs with some admixture of WDs and unevenly
distributed ESD binaries. The separation of the USDs from the rest
stars is explained by some differences in their spectra. It allows
us to separate the USDs with some level of probability by use of
all the following empirical conditions:
 $$
 \displaylines{\hfill
 (J-Ks)>0.06\hfill\cr\hfill
 (B_{SC}-I_{SC})>0.75\hfill\cr\hfill
 (B_{SC}-R_{SC})>0.42\hfill\cr\hfill
 (R_{UCAC3}-Ks)>1.25\hfill\llap(8)\cr\hfill
 (B_{SC}-Ks)>1.8-(R_{UCAC3}-R_{SC})\hfill\cr\hfill
 (J-Ks)>0.15, \hfill
 }
$$
\label{separ}

\begin{figure*}
\begin{center}
\includegraphics[width = 140mm]   {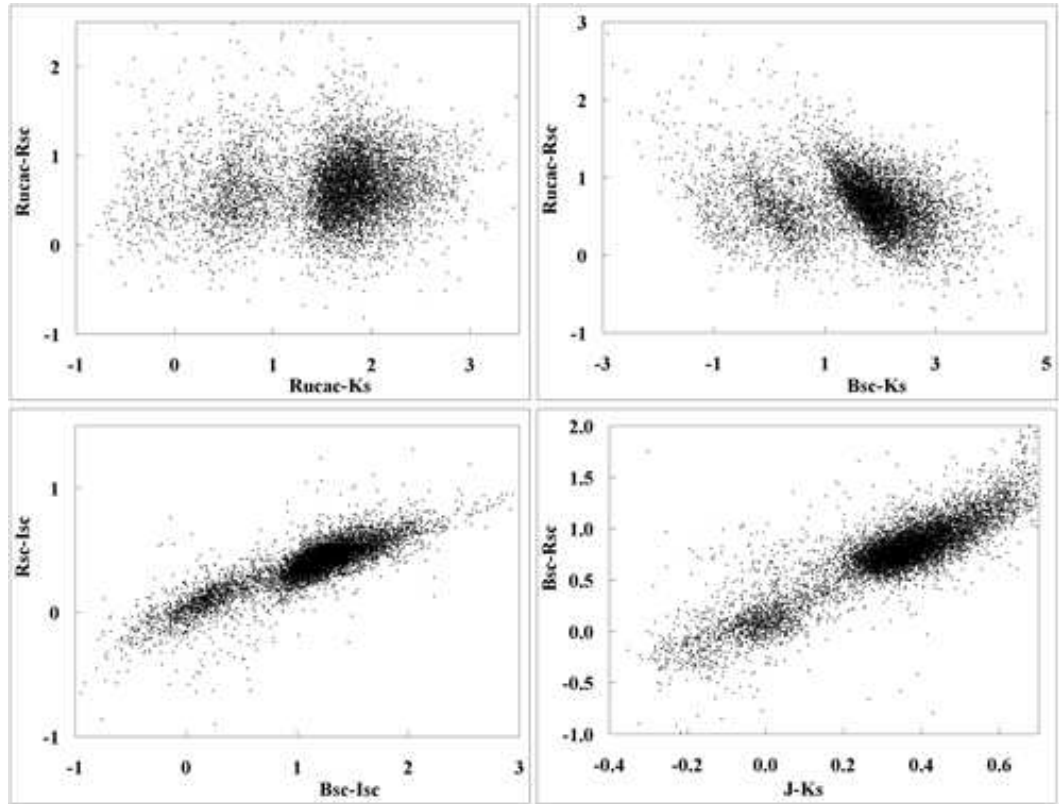}
\caption{Some color-color diagrams to separate USDs form the rest stars.}
\end{center}
\end{figure*}

The formula~\ref{poly} and \ref{separ} are used to separate 3 subsamples
among the 9799 selected stars: 7769 USDs, 1996 ESDs and 34 WDs.

\section{Statistics of the selected stars}

The distribution of the selected stars on $J$ magnitude is shown in Fig.~9:
dashed-dotted line for WDs, dotted one for ESDs, dashed one for USDs and solid line
for total distribution. Some overdensities at $J\approx7^{m}$,
$\approx10.5^{m}$ and $\approx13.5^{m}$
are related to the maxima in the Hipparcos, Tycho-2 and 2MASS distribution respectively.
It means that the selection criteria are stronger for fainter stars.
Consequently, the completeness decreases with magnitude and distance so that the ESD and USD
samples are almost complete to $J\approx8^{m}$ or distance $R\approx60$ pc and substantially
($\approx70\%$) complete to $J\approx10.5^{m}$ or distance $R\approx200$ pc.

\begin{figure}
\begin{center}
\includegraphics[width = 84mm]   {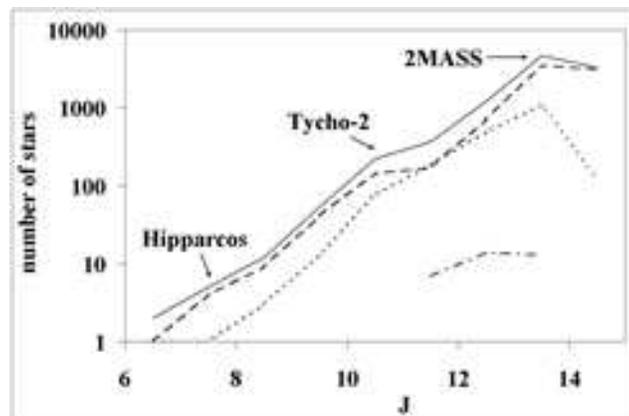}
\caption{The distribution of the selected stars on $J$ magnitude:
dashed-dotted line for WDs, dotted one for ESDs, dashed one for USDs and solid line
for total distribution. Some overdensities due to catalogue maxima are marked.}
\end{center}
\end{figure}

Hipparos contains 153 selected stars. There is spectral
classification for 140 stars from Tycho Spectral Types catalog
(TST) by \cite{tst} or other sources. In many cases the
classification is marked as doubtful. Only 27 stars are classified
as SDs and 10 as WDs.

\begin{figure*}
\begin{center}
\includegraphics[width = 140mm]   {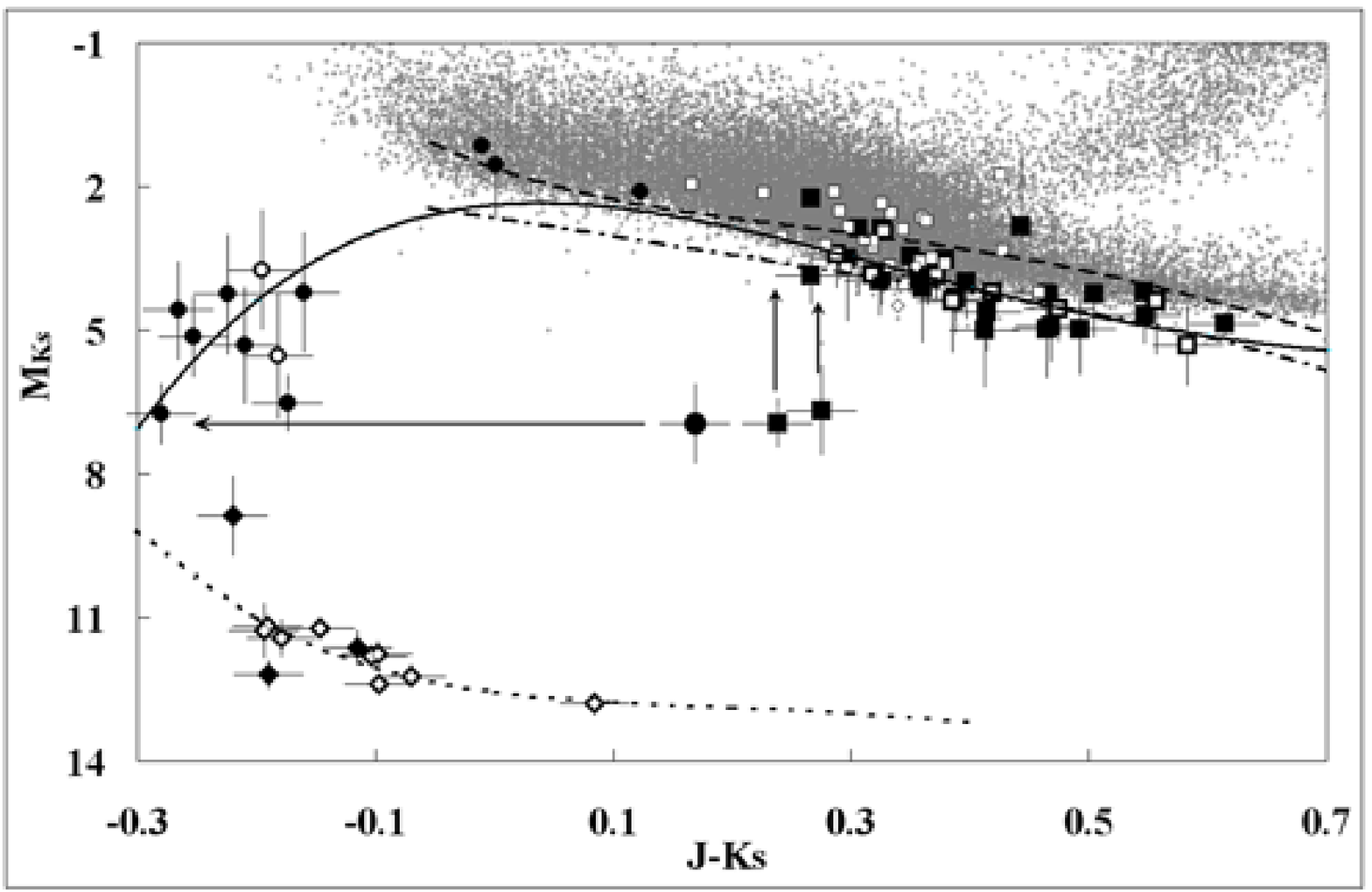}
\caption{Hipparcos stars with parallax relative error less than 0.3 in
the diagram ``$(J-Ks)$ vs. $M_{Ks}$'': all stars (grey ``cloud'' of points),
selected ESDs (black and open bold circles),
selected WDs (black and open bold diamonds), selected USDs (black and open bold squares),
the same categories with spectroscopic classification from the TST and
other sources (the same but bold and light open signs).
The dash-dotted and dashed lines show the isochrones for metallicities $Z=0.001$ and $Z=0.019$.
The accepted calibrations are shown for SDs (solid line) and WDs (dotted line).
The arrows show suspected replacement of 3 discussed outliers.}
\end{center}
\end{figure*}

The Fig.~10 presents the Hipparcos stars with parallax relative error less than 0.3 in
the diagram ``$(J-Ks)$ vs. $M_{Ks}$'' where $M_{Ks}$ is calculated from the Hipparcos parallaxes:
all stars (grey ``cloud'' of points),
selected ESDs (black and open bold circles), selected WDs (black and open bold diamonds),
selected USDs (black and open bold squares),
the same categories with spectroscopic classification from the TST and
other sources (the same but open light and open bold signs).
In the figure the open signs of classified stars are drawn over black ones of the selected
stars.
Therefore, finally the black signs are the selected stars without right spectral classification,
the open bold signs are the cases when our identification fits spectral classification
(2 ESDs, 9 WDs and 10 USDs)
and the light open signs are the stars which are classified but not selected
(1 ESD, 3 WDs and 23 USDs).
It is evident from the absolute magnitudes that the ESD and WDs which are classified but not
selected have wrong spectral classification.
Perhaps, the same is right for some of the classified but not selected USDs.
Thus, the mistakes in spectral classification of the subluminous stars are common.
Our method helps to verify the status of the stars.
The comparison of the fainter (non-Hipparcos) stars of our samples with the known
lists of classified subluminous stars will be provided elsewhere.

Three selected stars with $0.1<(J-Ks)<0.3$ and $M_{Ks}\approx7$ are interesting examples
of the complexity of correct selection of subluminous stars. They are
HIP~47296, 10529 and 3446.
HIP~47296 has been classified by spectrum as a suspected white dwarf.
Its ``$(J-Ks)$ -- $(B_{T}-V_{T})$'' relation proposes duplicity and composite spectrum.
Probably its duplicity also effects parallax giving its large error of 3 mas.
But no way to fit the parallax and $M_{Ks}$ for a WD.
$(B_{T}-V_{T})=-0.53$ in combination with $M_{Ks}=7$ suspects a SD component instead of WD.
For this star we accept the Hipparcos $M_{Ks}$ instead of the calibrated one.
HIP~10529 is a known pair with HIP~10531, probably optical.
Its parallax may be wrong. The colors are agreed, therefore, it must be a USD.
HIP~3446 has large error of the parallax (3 mas) and a discrepancy of the colors.
It must mean duplicity with a component to be a USD.
The arrows in Fig.~10 show the suspected replacement of these 3 outliers to their real positions
in the plane.
We conclude that duplicity is very important for the treatment of the
subluminous stars.

The lines in the Fig.~10 are related to the ``$(J-Ks)$ vs.
$M_{Ks}$'' calibration. The dotted line shows the accepted
empirical calibration for WDs close to the theoretical calibration
for the TRILEGAL code by \cite{g2005}:
 \begin{equation}
 M_{Ks}=-23.364(J-Ks)^{4}+33.36(J-Ks)^{3}-14.051(J-Ks)^{2}+3.346(J-Ks)+12.53
\label{wdcal}
 \end{equation}
Its accuracy due to intrinsic spread of the data is 0.2$^{m}$.

The dash-dotted and dashed lines show the isochrones for
metallicities $Z=0.001$ (typical for thick disk and halo) and
$Z=0.019$ (solar one) respectively \cite{g2000}. The former fits
redder USDs whereas the latter does bluer ones and some ESDs.
Generally it fits the suspects about their metallicities and ages.
The solid line is the accepted polynomial calibration
 \begin{equation}
M_{Ks}=18.28(J-Ks)^{4}-42.968(J-Ks)^{3}+30.359(J-Ks)^{2}-2.1323(J-Ks)+2.38
\label{sdcal}
 \end{equation}
Its intrinsic spread is about 0.3$^{m}$ for USDs and about 1$^{m}$
for ESDs. In fact it reflects the homo/heterogeneous nature of the
subsamples.

It has been shown that the $(J-Ks)$ for some binaries can be erroneous. Therefore, for all
the selected stars with color index discrepancies the empirical calibration is used:
\begin{equation}
M_{Ks}=0.5442M'_{Ks}-0.876
\label{mm}
\end{equation}
The accuracy of this calibration is estimated from the spread of the Hipparcos stars with
well-known parallaxes as about 1$^{m}$.

Photoastrometric ($R_{RPM}$, in cases of color index discrepancies) and
photometric ($R_{ph}$, in rest cases) distances as well as related rectangular coordinates XYZ
are calculated for the selected stars by use of the calibrated $M_{Ks}$
from $\lg(R_{ph})=(Ks-M_{Ks}+5-A_{Ks})/5$
(for the sake of simplicity all the obtained distances are referred as $R_{ph}$ hereafter).
The interstellar extinction $A_{Ks}$ generally no more than 0.2$^{m}$ for the $R<2$ kpc and,
hence, lower than the errors of the photometry and calibration. Therefore, the extinction
is ignored.

\begin{figure}
\begin{center}
\includegraphics[width = 84mm]   {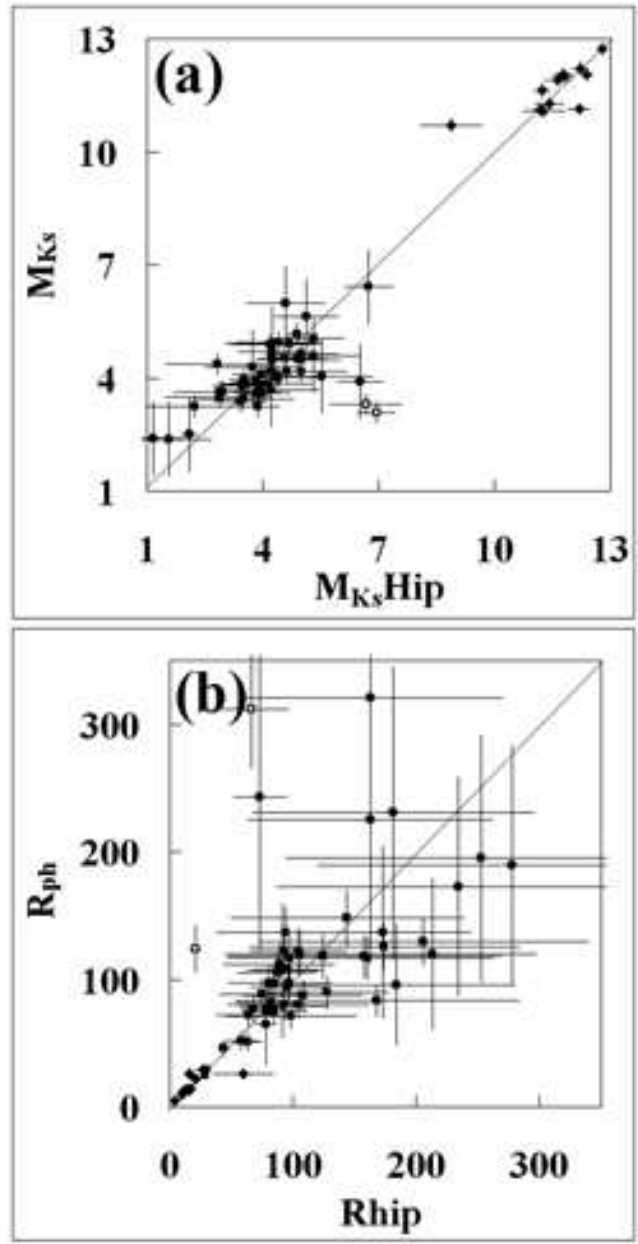}
\caption{The relations (a) $M_{Ks}$ vs. $M_{Ks}Hip$ and (b) $R_{ph}$ vs. $R_{HIP}$ (in pc)
for WDs (diamonds), ESDs (circles) and USDs (squares) with precise parallaxes.
The mentioned binaries are open signs.}
\end{center}
\end{figure}

The relations between the obtained and Hipparcos values are shown in Fig.~11:
(a) $M_{Ks}$ vs. $M_{Ks}HIP$, (b) $R_{ph}$ vs. $R_{HIP}$ (in pc). The WDs, ESDs and USDs with precise
parallaxes (relative error less than 0.3) are shown
by diamonds, circles and squares respectively.
Two mentioned binaries (HIP~10529 and 3446) are shown by open signs.
Unfortunately, for ESDs the both distances, $R_{ph}$ and $R_{HIP}$ have low accuracy.

The distribution of all the selected stars on the celestial sphere
in galactic coordinates is shown in Fig.~12: (a) ESDs, (b) USDs
and (c) WDs. The same for the Tycho-2 selected stars is shown in
Fig.~12 (d), (e), (f). It is seen that no WDs at high latitudes.
The distribution of all USDs is strongly affected by the UCAC3
selection in favor to the southern equatorial hemisphere. The
distributions of the ESDs and USDs show some voids at the galactic
centre and along the Gould belt because of the extinction
considered by \cite{gould}. Despite the voids both ESDs and USDs
show considerable concentration to the galactic centre hemisphere.

Some overdensities of the ESDs are seen both among Tycho-2 and UCAC3-XPM stars at:
\begin{enumerate}
\item $l\approx218$, $b\approx+5$,
\item $l\approx278$, $b\approx-32$, in front of the LMC,
\item $l\approx287$, $b\approx-2$, at $\eta~Car$ region,
\item $l\approx314$, $b\approx+15$, at Sco-Cen association,
\item $l\approx318$, $b\approx-12$,
\item $l\approx332$, $b\approx-2$,
\item $l\approx5$, $b\approx-42$,
\item $l\approx22$, $b\approx-32$,
\item $l\approx33$, $b\approx-41$,
\item $l\approx94$, $b\approx-2$,
\item $l\approx137$, $b\approx-22$
\end{enumerate}
A detail investigation of individual stars in the overdensities shows that they must be real.
The nature of the overdensities is to be revealed.
One may suspect them as some remnants of dwarf galaxies or globular clusters.
Else, since some of the overdensities are near star formation regions, one could suspect
an intense mass loss for the stars near the regions making so many ESDs.

\begin{figure*}
\begin{center}
\includegraphics[width = 140mm]   {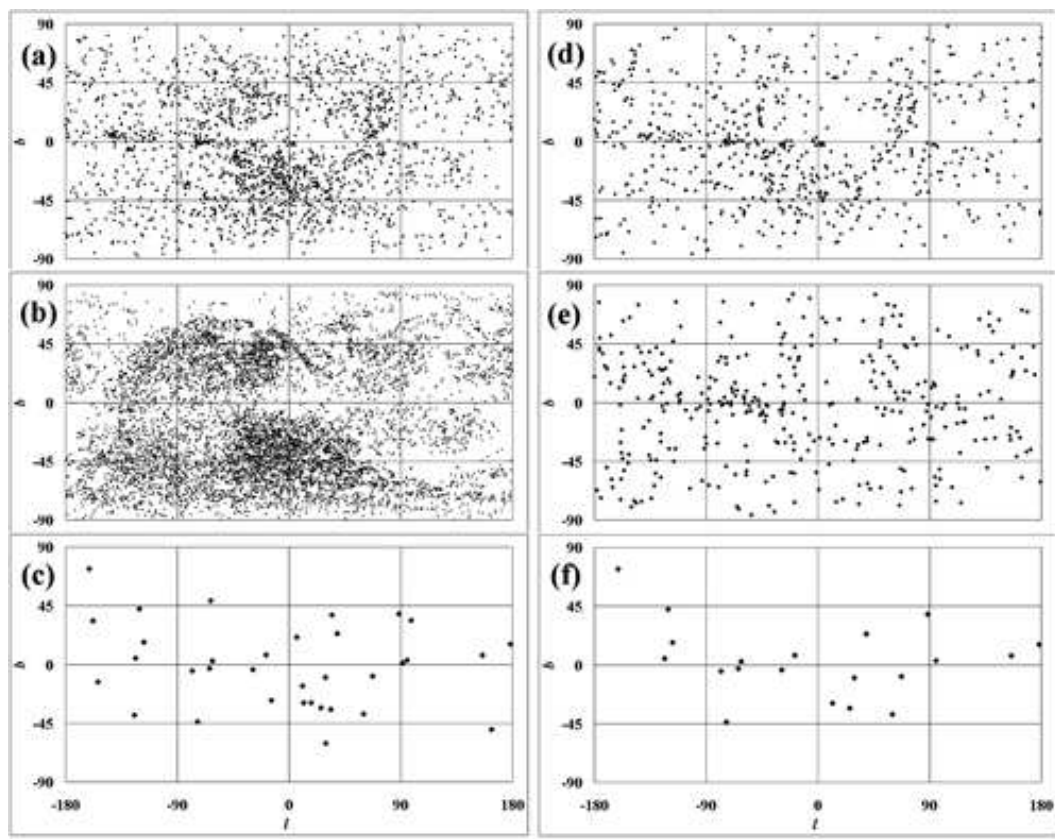}
\caption{The distribution of all selected stars on the celestial sphere in galactic coordinates:
(a) ESDs, (b) USDs and (c) WDs and the same for the Tycho-2 selected stars: (d), (e), (f).}
\end{center}
\end{figure*}

\section{3D distribution and motion of the stars}

The distribution of the selected ESDs is shown in Fig.~13:
all ESDs projected into the (a) XY, (b) XZ and (c) YZ planes as well as
Tycho-2 ESDs projected into the same planes in (d), (e) and (f) subfigures
(the distances in kpc).
The same data for USDs are shown in Fig.~14.
The distribution of the selected WDs is not shown because all of them are within 50 pc.
The voids by the extinction near the galactic plane and in the Gould belt are evident
in Fig.~14 for all subsamples quite dense at least to 500 pc.
The Gould belt voids are marked by arrows in some XZ subfigures.
Higher stellar density in the galactic centre hemisphere is evident.
It gives new constraints to galactic models and will be analyzed in detail in next papers.

\begin{figure}
\begin{center}
\includegraphics[width = 84mm]   {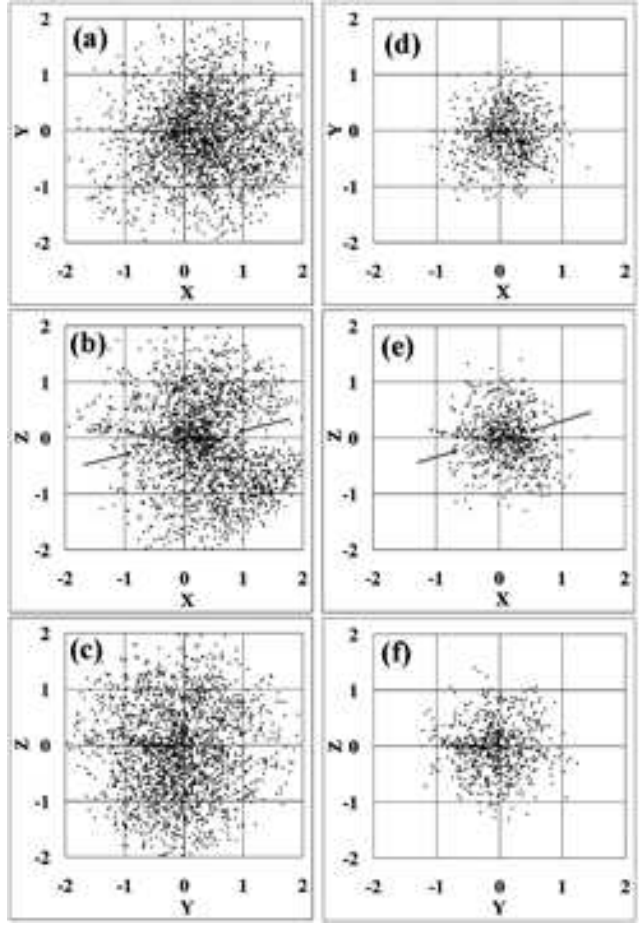}
\caption{The distribution of all selected ESDs projected into the (a) XY, (b) XZ and
(c) YZ planes as well as Tycho-2 ESDs projected into the same planes in (d), (e) and (f)
subfigures (the distances in kpc).}
\end{center}
\end{figure}

\begin{figure}
\begin{center}
\includegraphics[width = 84mm]   {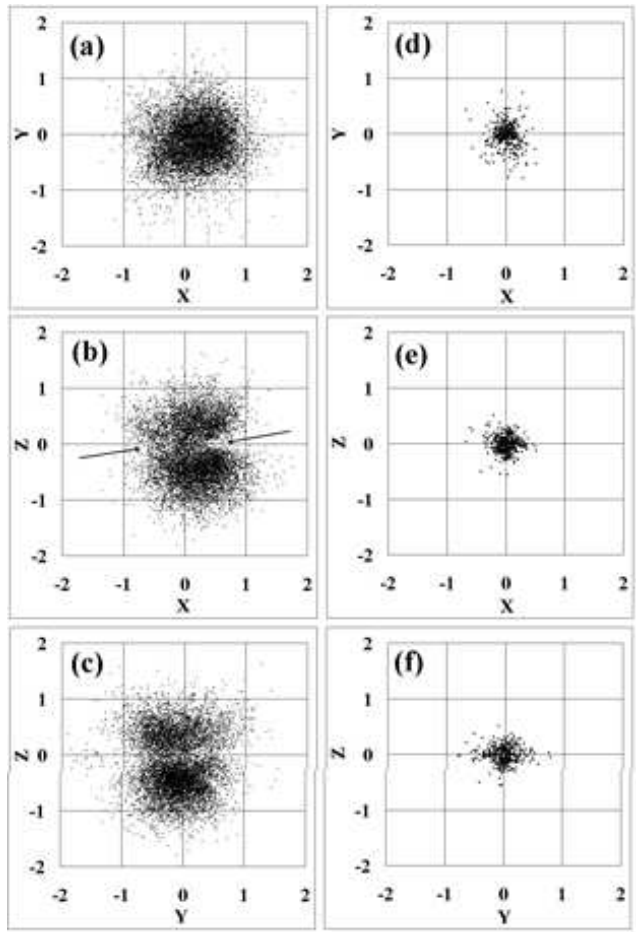}
\caption{The distribution of all selected USDs projected into the (a) XY, (b) XZ and
(c) YZ planes as well as Tycho-2 USDs projected into the same planes in (d), (e) and (f)
subfigures (the distances in kpc).}
\end{center}
\end{figure}

The Pulkovo Compilation of Radial Velocities (PCRV) catalogue
\cite{pcrv} contains radial velocities (RVs) for 78 selected stars
with precision better than 5 km~s$^{-1}$. Less precise RVs are
collected from other sources for more 105 selected stars. Precise
RVs are the crucial data for modern kinematic analysis of any
stellar sample. Some advances in RV data are expected in near
future. Therefore, in this {\em preliminary} study of the sample
kinematics we use either 78 best RVs or all 183 RVs.

These RVs are used
together with $\alpha$, $\delta$, $\mu$ and $\pi$ or $R_{ph}$ to calculate rectangular
components U, V, W of the stellar space motion with respect to the Sun.
No correction for galactic rotation and solar motion to the apex applied.

The Hipparcos distances highly correlate with $R_{ph}$ for 78 best stars. Therefore,
the main features of the stellar distribution in the 6D space of XYZUVW is the same
for the both sets based on Hipparcos distances or $R_{ph}$.
Here we consider only XYZUVW based on $R_{ph}$.

The mean and dispersion for U, V and W velocity components for 11
ESDs and 66 USDs is presented in Table~2 (in km~s$^{-1}$). The
asymmetric drift (considerable negative $\overline{V}$) is evident
for both ESDs and USDs. For USDs it is similar to the Besan\c{c}on
galactic model value for halo: -226 km~s$^{-1}$ \cite{robin}. For
ESDs it is between the Besan\c{c}on halo and thick disk (-53
km~s$^{-1}$) values proving that the ESDs is a heterogeneous
sample with disk and halo stars. The same conclusions follow from
the dispersions: the USDs show the dispersions similar to
Besan\c{c}on halo (131, 106, 85 km~s$^{-1}$ for U, V, W) biased
due to selection in favor to faster stars whereas the ESD
dispersions fall between the Besan\c{c}on halo and thick disk (67,
51, 42 km~s$^{-1}$ for U, V, W).

\begin{table}
 \caption{The mean/dispersion for U, V and W velocity components for ESDs and USDs (in km~s$^{-1}$).}
 \label{uvw}
 \begin{tabular}{@{}lccc}
 \hline
  & U & V & W  \\
\hline
ESD    & 30/92 & -124/113 & 5/69 \\
USD    & -23/189 & -267/94 & -27/110 \\
\hline
 \end{tabular}
 \medskip
\end{table}

The distribution of 1 WD (diamond), 11 ESDs (circles) and 66 USDs (open squares)
with precise RVs in the projection to the UV, UW and VW planes (velocities in km~s$^{-1}$)
is shown in Fig.~15. The dispersion of the velocities and asymmetric drift in the V component
are clearly seen allowing kinematic classification of the stars.
The WD and most ESDs certainly belong to the galactic thin or thick disk (population I).
As expected, most USDs belong to the halo (population II). Some of the USDs have retrograde
motion.
However, the distribution of the USDs along U and V is not normal. It means a strong bias due to
the selection in favor to the higher velocities with respect to the Sun (as mentioned above,
many slow SDs are lost due to selection method).

\begin{figure}
\begin{center}
\includegraphics[width = 70mm]   {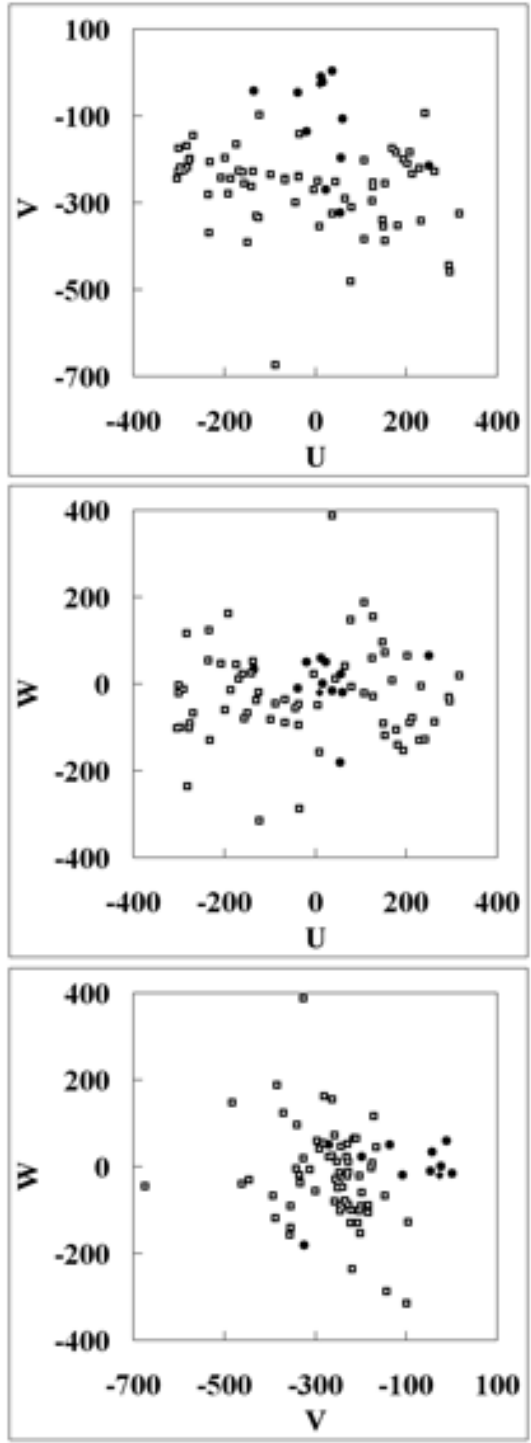}
\caption{The distribution of WD (diamond), ESDs (circles) and USDs (open squares)
with precise RVs in the projection to the UV, UW and VW planes (velocities in km~s$^{-1}$).}
\end{center}
\end{figure}

Metallicities Fe/H are collected for 56 stars from various
sources. The relation of the Fe/H and velocity component V is
shown in Fig.~16 for ESDs (circles) and USDs (open squares). The
error bars show the estimation of the accuracy of V from the ones
of $\mu$, RV and $R_{ph}$ as well as the error of about 0.3 dex
accepted for Fe/H. The solar position is marked. The vertical
lines show the separation of thin/thick disk at about
Fe/H$\approx-0.3$ (Z$\approx0.01$) and thick disk/halo at about
Fe/H$\approx-1.3$ (Z$\approx0.001$) similar to \cite{robin}. The
ESDs are found both in halo and disk. The thin-thick disk
separation is not evident in these data. The USDs belong to halo
and probably to thick disk (taking into account that the errors
for the most metal rich USDs are at the halo/disk boundary). The
velocity-metallicity trend is evident for the USDs. For the ESDs
only 2 stars in the halo domain are velocity outliers. The rest
ESDs show no trend. One of 2 high velocity low metallicity ESD
outliers, HIP~103755 is known RR~Lyr variable \cite{hip}.

\begin{figure}
\begin{center}
\includegraphics[width = 84mm]   {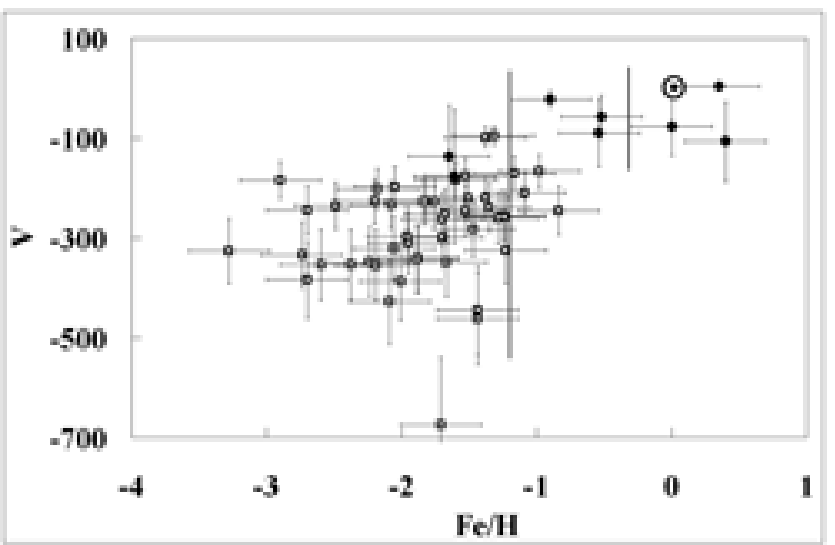}
\caption{The relation of the Fe/H and velocity component V for ESDs (circles) and USDs (open squares).}
\end{center}
\end{figure}

There is a relation between Fe/H of SD and its position in the
``$(J-Ks)$ vs. $M_{Ks}$'' diagram. The ESDs (circles) and USDs
(open squares) are shown in Fig.~17 in the diagram (a) ``$M_{Ks}$
vs. Fe/H'', (b) ``$(J-Ks)$ vs. Fe/H''. The relation for ESDs is
due to 2 most luminous stars mentioned above with high velocity
and low metallicity. The relations for the USDs can be explained
by ``$(J-Ks)$ -- age'' and ``$(J-Ks)$ -- metallicity'' relations
\cite{g2000}: the turn-off point becomes redder with rise of the
age and metallicity. Therefore, one can see that the redder part
of the USD subsample contains old stars with moderate metallicity
whereas the bluer part has larger dispersion of the Fe/H because
of the admixtures of younger stars with higher Fe/H as well as
older stars with lower Fe/H.

\begin{figure}
\begin{center}
\includegraphics[width = 84mm]   {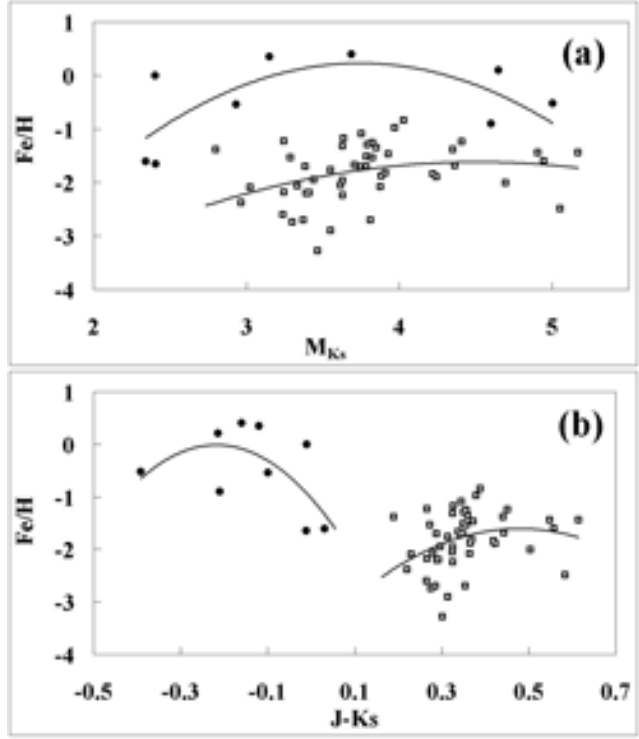}
\caption{The relation of the Fe/H and (a) $M_{Ks}$, (b) $(J-Ks)$ for ESDs (circles)
and USDs (open squares).}
\end{center}
\end{figure}

The galactic orbits are calculated for 183 selected stars with
XYZUVW set accepting the solar galactocentric distance of 8.5 kpc.
The model of the gravitational potential of the Galaxy by
\cite{as} is adopted. The orbits were calculated over a total of
1.1 Gyr backwards.

The eccentricities for 3 WDs are 0.05 for 2MASS PSC 1098632336, 0.12 for HIP~14754 and 0.36 for HIP~101516.
The projections of the WD orbits into XY, XZ and YZ planes are shown in Fig.~18, row (a) by solid, dotted
and dashed curves for different stars (distances in kpc).
All of these orbits better fit to the thin disk.

The mean eccentricity for 41 ESDs is $0.46\pm0.3$ and the one for 139 USDs is $0.8\pm0.2$.
The projections of the ESD and USD orbits into XY, XZ, YZ planes are shown in Fig.~18, rows (b) and (c)
respectively (distances in kpc). One should pay attention to the different scales in the subfigures.

\begin{figure}
\begin{center}
\includegraphics[width = 140mm]   {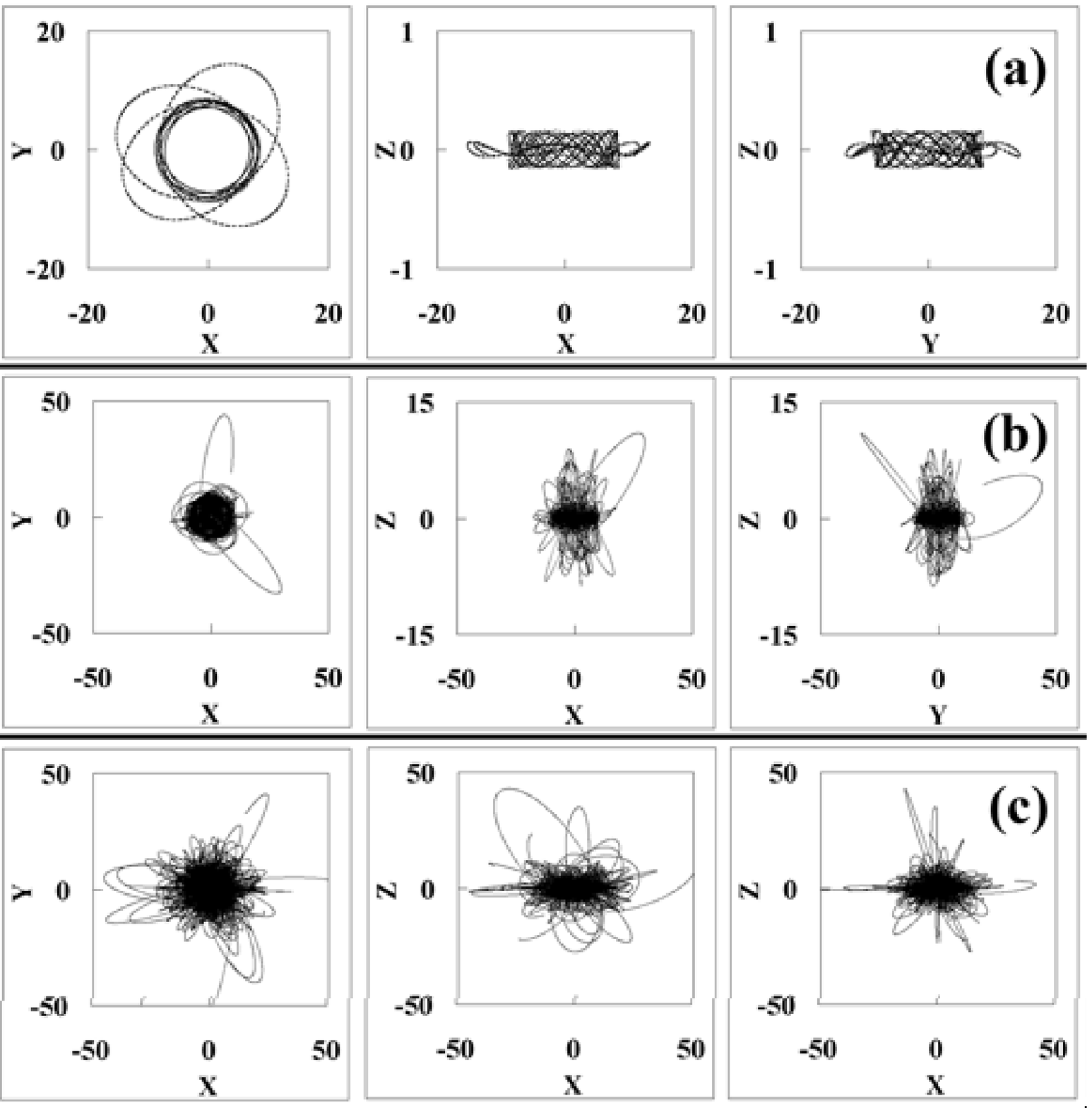}
\caption{The projections of the (a) WD, (b) ESD and (c) USD orbits into XY, XZ and YZ planes.
The scales are different.}
\end{center}
\end{figure}

The distributions of ESDs (dashed line) and USDs (solid line) on the orbital eccentricity and
apogalactic distance Zmax are shown in Fig.~19 (a) and (b) respectively.
The vertical lines show suspected separation into thin disk, thick disk and halo.

In Fig.~20 we show the average spatial distribution of ESDs
(dashed line) and USDs (solid line) on Z distance (in kpc) based
on their orbits. It is given by the statistics of the orbital
points calculated with equal time steps of 1 Myr. It is similar to
the analysis of the galactic orbits of sdB stellar sample by
\cite{boer}. In contrast to them, our results expand far beyond
$|Z|=4$ kpc and our distributions do not have local minima at
Z=$0$. The reason of the difference is that their choice of stars
was solely determined by the availability of the data giving some
selection. But our sample is deeper, and near the Sun it is almost
complete without any selection. Within $|Z|<2.5$ kpc both ESD and
USD samples are dominated by the thick disk stars. Their spatial
distribution in Z is fairly the same for ESDs and USDs and fits an
exponential distribution with a scale height of about $1.25\pm0.1$
kpc. The wings of the distributions are populated by the halo ESD
and USD stars. The halo ESDs are rare and show no distribution
low. The USD halo spatial distribution in Z fits an exponential
one with a scale height of about $8\pm1$ kpc. Our USD halo is so
well-defined that we can calculate its local mass density assuming
mean stellar mass of $0.5M_\odot$: $\rho_{0}=2\cdot10^{-5}
M_\odot~pc^{-3}$. It is nearly 2 times higher than the local mass
density for all halo star accepted for the Besan\c{c}on model of
the Galaxy by \cite{robin}, although the Besan\c{c}on value is
initial mass function dependent. This density increase may be due
to our sample depth and completeness.

The relations of Fe/H versus eccentricity, perigalactic distance Rmin (in kpc) and
apogalactic distance Rmax (in kpc) are shown in Fig.~21 (a), (b) and (c) respectively
for ESDs (circles) and USDs (open squares).
It is evident that 2 ESD outliers mentioned above have halo member properties similar to the USDs.
The rest ESDs show the properties of disk members.

The conclusions from the Fig.~19, 20 and 21 are the same:
most ESDs belong to thick disk whereas most USDs belong to halo.

\begin{figure}
\begin{center}
\includegraphics[width = 84mm]   {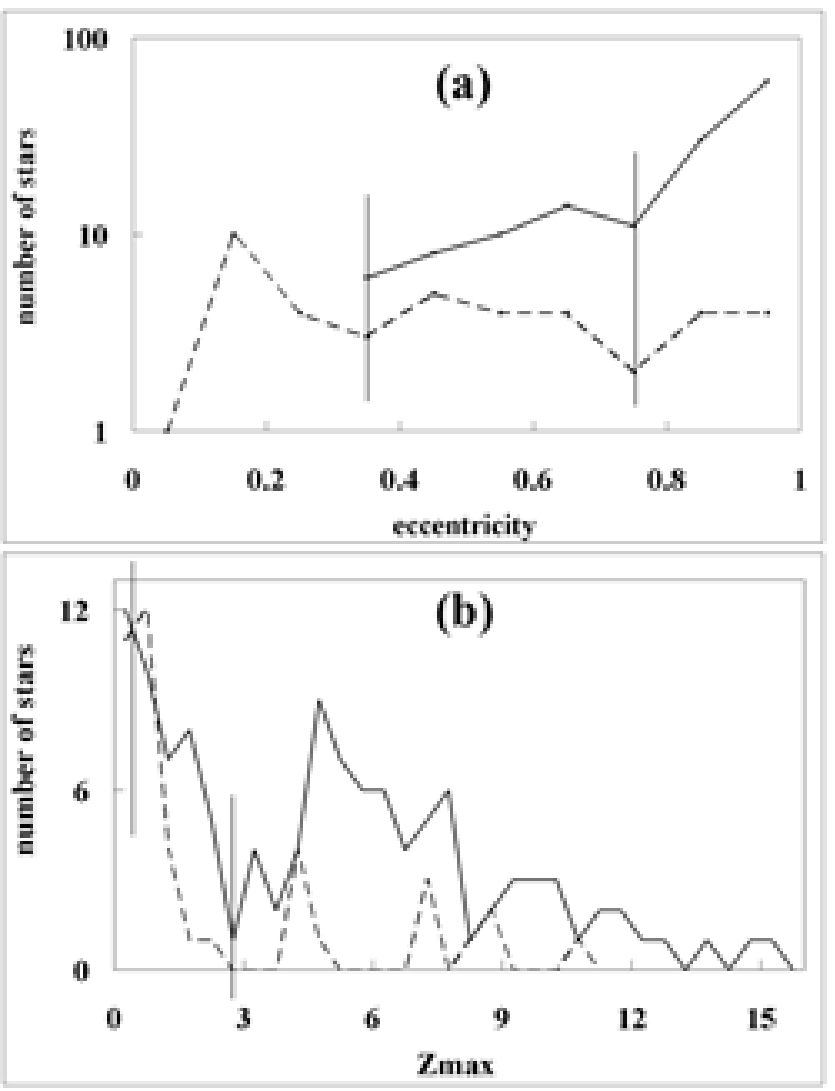}
\caption{The distribution of ESDs (dashed line) and USDs (solid line) (a) on orbital eccentricity
and (b) on apogalactic distance Zmax.
The vertical lines show suspected separation into thin disk, thick disk and halo.}
\end{center}
\end{figure}

\begin{figure}
\begin{center}
\includegraphics[width = 84mm]   {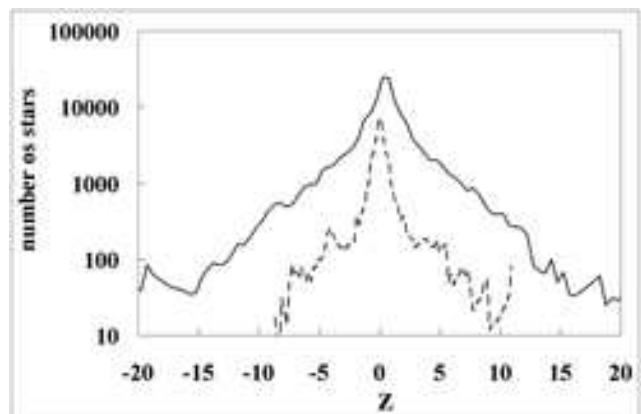}
\caption{The average spatial distribution of ESDs (dashed line) and USDs (solid line) on Z
distance (in kpc) based on their orbits.}
\end{center}
\end{figure}

\begin{figure}
\begin{center}
\includegraphics[width = 84mm]   {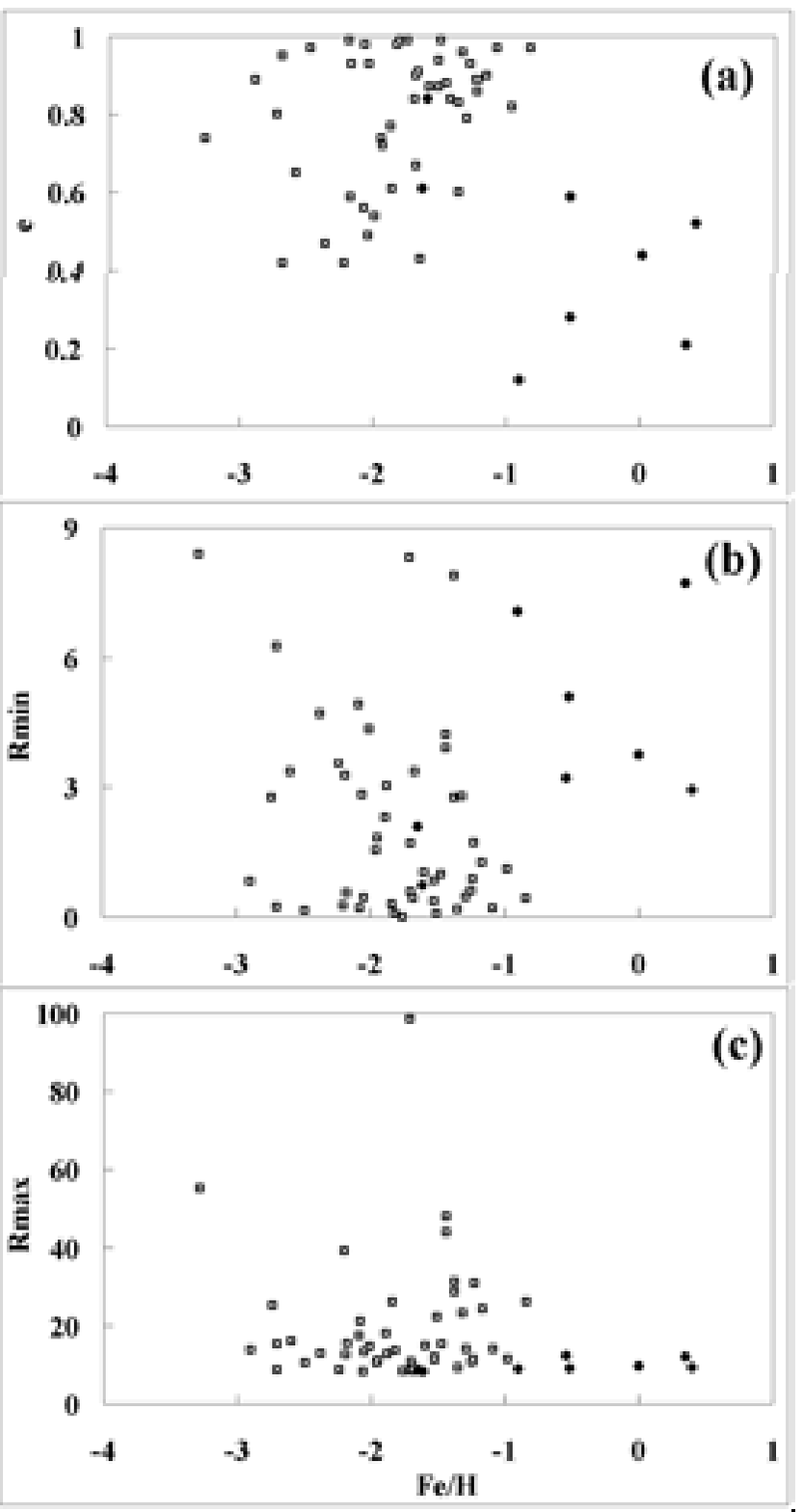}
\caption{The relations of Fe/H versus eccentricity, perigalactic and
apogalactic distance (in kpc) for ESDs (circles) and USDs (open squares).}
\end{center}
\end{figure}

\section{Catalogue format}

The properties of the 9799 selected stars are presented in electronic form as the catalogue
``Subdwarfs and white dwarfs from the 2MASS, Tycho-2, XPM and UCAC3 catalogues''
(hereafter SDWD catalogue).
The table~\ref{cata} gives its format.
The catalogue data are arranged by the stellar class from 1 (WD) to 2 (ESD) and 3 (USD)
as well as obtained $M_{Ks}$ increasing within the class).

\begin{table}
 \caption{Format of the catalogue of subdwarfs and white dwarfs from the 2MASS,
Tycho-2, XPM and UCAC3.}
 \label{cata}
 \begin{tabular}{@{}llc}
 \hline
field & format & positions \\
\hline
2MASS psc number                    &   I10  &   1-10   \\
Tycho-2 number                      &   A10  &  11-20   \\
Hipparcos number                    &   I6   &  21-26   \\
stellar class (WD=1, ESD=2, USD=3)  &   I1   &  27-27   \\
obtained $M_{Ks}$                   &   F5.2 &  28-32   \\
$l$ in decimal degs                 &   F8.4 &  33-40   \\
$b$ in decimal degs                 &   F8.4 &  41-48   \\
$\mu_{l}\cos(b)$ in mas~yr$^{-1}$   &   F7.1 &  49-55   \\
$\mu_{b}$ in mas~yr$^{-1}$          &   F7.1 &  56-62   \\
Hipparcos parallax in mas           &   F6.2 &  63-68   \\
parallax precision in mas           &   F4.1 &  69-72   \\
$B_{T}$ mag from Tycho-2            &   F5.2 &  73-77   \\
$B_{T}$ precision from Tycho-2      &   F4.2 &  78-81   \\
$V_{T}$ mag from Tycho-2            &   F5.2 &  82-86   \\
$V_{T}$ precision from Tycho-2      &   F4.2 &  87-90   \\
$J$ mag from 2MASS                  &   F5.2 &  91-95   \\
$Ks$ mag from 2MASS                 &   F5.2 &  96-100  \\
$R_{UCAC3}$ mag from UCAC3          &   F5.2 & 101-105  \\
$R_{UCAC3}$ precision from UCAC3    &   F4.2 & 106-109  \\
$B_{SC}$ mag from SuperCosmos       &   F5.2 & 110-114  \\
$R_{SC}$ mag from SuperCosmos       &   F5.2 & 115-119  \\
$I_{SC}$ mag from SuperCosmos       &   F5.2 & 120-124  \\
$B_{SC}$ quality flag from UCAC3    &   I1   & 125-125  \\
$R_{SC}$ quality flag from UCAC3    &   I1   & 126-126  \\
$I_{SC}$ quality flag from UCAC3    &   I1   & 127-127  \\
obtained photometric distance in pc &   I4   & 128-131  \\
obtained X distance in pc           &   I5   & 132-136  \\
obtained Y distance in pc           &   I5   & 137-141  \\
obtained Z distance in pc           &   I5   & 141-146  \\
radial velocity in km~s$^{-1}$      &   F6.1 & 147-152  \\
RV precision in km~s$^{-1}$         &   I2   & 153-154  \\
Fe/H                                &   F4.1 & 155-158  \\
epoch difference from XPM in years  &   F5.2 & 159-163  \\
velocity component U in km~s$^{-1}$ &   F6.1 & 164-169  \\
velocity component V in km~s$^{-1}$ &   F6.1 & 170-175  \\
velocity component W in km~s$^{-1}$ &   F6.1 & 176-181  \\
apogalactic distance in kpc         &   F4.1 & 182-185  \\
perigalactic distance in kpc        &   F3.1 & 186-188  \\
galactic orbit eccentricity         &   F4.2 & 189-192  \\
\hline
 \end{tabular}
 \medskip
\end{table}

\section{Conclusions}

New all-sky astrometric and photometric surveys (2MASS, XPM, UCAC3, SuperCosmos, Tycho-2)
appear suitable to make the largest sample of subluminous stars candidates:
7769 unevolved subdwarfs, 1996 evolved subdwarfs and 34 white dwarfs.
The key feature of this selection is the detailed analysis of the distribution of all stars
from these catalogues in the various ``color index vs. reduced proper motion'' planes
as well as Monte-Carlo simulation of this distribution.
It is found that only surveys with proper motion accuracy better than 10 mas~yr$^{-1}$
provide the acceptable separation of the subluminous stars from the main sequence, while
with 10\% admixture and considerable selection biases in favor to faster stars.
It is proved that most of the Tycho-2, XPM and UCAC3 stars fit this criterion of the proper
motion accuracy.
It is pointed out that future surveys with proper motion accuracy better than 1 mas~yr$^{-1}$
can provide a perfect separation without an admixture and bias.
The multi-color photometry also appears useful for the subluminous star selection and separation
of evolved and unevolved subdwarfs as well as single and binary stars.
It is pointed out that current level of the photometric accuracy is acceptable for such
tasks but the level of about 0.01$^{m}$ is desirable for better separation.

The calibrations ``color index vs. absolute magnitude'' and ``reduced proper motion vs. absolute magnitude''
made with the best Hipparcos stars allow us to calculate photometric and photoastrometric distances
for all the selected stars, consider their 3D distribution, while complete only near the Sun.
The use of radial velocities
for 183 stars and Fe/H metallicities for 56 stars allow us to consider their 3D motion and
metallicity-velocity relation.

This investigation tests some theories related to the subluminous
stars. Namely, it is shown that unevolved and evolved subdwarfs
have different evolutionary status, kinematics and metallicity.
The most USDs are population II low metallicity high asymmetric
drift stars from halo with the scale height of $8\pm1$ kpc and
local mass density of the halo USDs of $2\cdot10^{-5}
M_\odot~pc^{-3}$. The ESDs are a heterogeneous group containing
stars from disk and halo with metallicities from low to solar one
and with halo to thin disk kinematics. However, most evolved
subdwarfs seems to belong to thick disk with scale height of
$1.25\pm0.1$ kpc. The evolved subdwarfs appear the most
interesting group. They show some spatial overdensities of yet
unknown nature, large fraction of binaries and vast diversity of
properties not all of which have been explained by current
theories.

All important parameters of the selected stars are compiled into the SDWD catalogue in order to
continue their investigations, specially to prove their status by spectroscopy.

\section{ACKNOWLEDGMENTS}

This study was supported by the Fundamental Researches State Fund of Ukraine
(project No. FRSF-28/238) and the Russian Foundation for Basic Research
(projects No. 08-02-00400 and No. 09-02-90443-Ukr-f), and in part by the ``Origin
and Evolution of Stars and Galaxies'' - ``Program of the Presidium of the Russian
Academy of Sciences and the Program for State Support of Leading Scientific Schools
of Russia'' (NSh-6110.2008.2).


\begin{thebibliography}{9}

\small


\bibitem [Allen \& Santill\'an, 1991]{as} Allen C., Santill\'an A., 1991, Rev. Mexicana Astron.
Astrof., 22, 255
\bibitem[Bertelli et al., 2008]{bgmn} Bertelli G., Girardi L., Marigo P.,
Nasi E., 2008, A\&A, 484, 815
\bibitem[de Boer et al., 1997]{boer} de Boer K.S., Aguilar Sanchez Y.,
Altmann M., Geffert M., Odenkirchen M., Schmidt J.H.K., Colin J., 1997, A\&A, 327, 577
\bibitem[Cassisi et al., 2003]{cssw} Cassisi, S., Schlattl G., Salaris M.,
Weiss A., 2003, APJ, 582, L43
\bibitem[Catelan, 2007]{catelan} Catelan, M., 2007, American Institute
of Physics Conf. Proc., 930, 39
\bibitem[Fagotto et al., 1994]{fbbc} Fagotto, F., Bressan A., Bertelli G.,
Chiosi C., 1994, A\&AS, 104, 365
\bibitem[Fedorov, Myznikov \& Akhmetov, 2009]{xpm} Fedorov P.N., Myznikov A.A.,
Akhmetov V.S., 2009, MNRAS, 393, 133
\bibitem[Fedorov et al., 2010]{xpm1} Fedorov P.N., Akhmetov V.S.,
Bobylev V.V., Bajkova A.T., 2010, MNRAS, in press
\bibitem[{Finlator et al., 2000}]{sdss} Finlator K., Ivezic Z., Fan X., et al.,
2000, AJ, 120, 2615
\bibitem[{Girardi et al., 2000}]{g2000} Girardi L., Bressan A., Bertelli G.,
Chiosi C., 2000, A\&AS, 141, 371
\bibitem[{Girardi et al., 2005}]{g2005} Girardi L., Groenewegen M.A.T.,
Hatziminaoglou E., da Costa L., 2005, A\&A, 436, 895
\bibitem[{Gontcharov, 2006}]{pcrv} Gontcharov G.A., 2006, Astronomy Letters, 32, 759
\bibitem[{Gontcharov, 2008a}]{ob} Gontcharov G.A., 2008a, Astronomy Letters,  34, 7
\bibitem[{Gontcharov, 2008b}]{rcg} Gontcharov G.A., 2008b, Astronomy Letters,  34, 868
\bibitem[{Gontcharov, 2009a}]{model} Gontcharov G.A., 2009a, Astronomy Letters, 35, 707
\bibitem[{Gontcharov, 2009b}]{gould} Gontcharov G.A., 2009b, Astronomy Letters, 35, 862
\bibitem[{Hambly, Irwin \& MacGillivray, 2001}]{sc} Hambly N.C., Irwin M.J.,
MacGillivray H.T., 2001, MNRAS, 326, 1295
\bibitem[Han et al., 2003]{han} Han Z., Podsiadlowski Ph., Maxted P.F.L.,
Marsh T.R., 2003, MNRAS, 341, 669
\bibitem[Hipparcos and Tycho catalogues, 1997]{hip} Hipparcos and Tycho catalogues,
1997. ESA
\bibitem[H\o{}g et al., 2000]{tycho2} H\o{}g E., Fabricius C., Makarov V.V.,
et al., 2000, A\&A, 355, L27
\bibitem[{Jones, 1972}]{jones} Jones E.M., 1972, APJ, 173, 671
\bibitem[{van Leeuwen, 2007}]{hip2} van Leeuwen F., 2007, A\&A, 474, 653
\bibitem[{Miller Bertolami et al., 2008}]{mb} Miller Bertolami M.M., Althaus L.G.,
Unglaub K., Weiss A., 2008, A\&A, 491, 253
\bibitem[{Monet, 1998}]{usnoa2} Monet D., 1998, BAAS, 30, 1427
\bibitem[{\O{}stensen, 2009}]{osten} \O{}stensen, 2009, in Schuh S. \& Handler G.,
eds., Proc. JENAM 2008 Symp. \#4, Asteroseismology and Stellar Evolution, 159, 75
\bibitem[{Robin et al., 2003}]{robin} Robin A.C., Reyle C., Derriere S.,
Picaud S., 2003, A\&A, 409, 523
\bibitem[{Skrutskie et al., 2006}]{2mass} Skrutskie M.F., Cutri R.M., Stiening R.,
et al., 2006, ApJ, 131, 1163
\bibitem[{Smith et al., 2009}]{smith} Smith M.C., Evans N.W., Belokurov V.,
et al., 2009, MNRAS, 399, 1223
\bibitem[{Stark \& Wade, 2003}]{stark} Stark M.A., Wade R.A., 2003, AJ, 126, 1455
\bibitem[{Vassiliadis \& Wood, 1994}]{vw} Vassiliadis E., Wood P.R., 1994, ApJS, 92, 125
\bibitem[{Veltz et al., 2008}]{veltz} Veltz L., Bienayme O., Freeman K.C.,
et al., 2008, A\&A, 480, 753
\bibitem[{Wichman \& Hill, 1982}]{random} Wichman B.A., Hill I.D., 1982,
Applied Statistics, 31, 188
\bibitem[{Wright et al., 2003}]{tst} Wright C.O., Egan M.P., Kraemer K.E.,
Price S.D., et al., 2003, AJ, 125, 359
\bibitem[{Zacharias et al., 2009}]{ucac3} Zacharias N., et al., 2009, AJ, in press




\end{thebibliography}
\end{document}